\documentclass[12pt]{article}
\usepackage{amsmath}
\usepackage{graphicx}
\usepackage{enumerate}
\usepackage{natbib}
\usepackage{url} 
\usepackage{binomexp}
\usepackage{mathtools}
\usepackage{amsfonts}
\usepackage{bbm}
\usepackage{dsfont}
\usepackage{longtable}
\usepackage{yhmath}
\usepackage{bm}
\usepackage{amssymb}
\usepackage[nodisplayskipstretch]{setspace}
\usepackage{tikz}
\usepackage{float}
\usepackage{color}
\usepackage{stmaryrd}
\usepackage{booktabs}
\usepackage{paralist}
\usepackage{multirow}
\usepackage{subfig}
\usepackage{threeparttable}
\usepackage[inline]{enumitem}
\usepackage{array}
\newcolumntype{P}[1]{>{\centering\arraybackslash}p{#1}}
\usepackage{algorithm,algpseudocode}
\makeatletter
\newcommand{\algmargin}{\the\ALG@thistlm}
\makeatother
\newlength{\whilewidth}
\algdef{SE}[parWHILE]{parWhile}{EndparWhile}[1]
  {\parbox[t]{\dimexpr\linewidth-\algmargin}{%
     \hangindent\whilewidth\strut\algorithmicwhile\ #1\ \algorithmicdo\strut}}{\algorithmicend\ \algorithmicwhile}%
\algnewcommand{\parState}[1]{\State%
  \parbox[t]{\dimexpr\linewidth-\algmargin}{\strut #1\strut}}

\newcommand{\bcx}{{\bm X}}

\newcommand{\bcv}{{\bm V}}

\newcommand{\bco}{{\bm O}}

\newcommand{\bcd}{{\bm D}}
\newcommand{\bcc}{{\bm C}}

\newcommand{\bbE}{\mathbb{E}}
\newcommand{\bbI}{\mathbb{I}}

\newcommand{\E}{\mathbb{E}}
\newcommand{\Prob}{\mathbb{P}}

\usepackage[colorlinks=true,linkcolor=red,citecolor=black]{hyperref}

\newtheorem{theorem}{Theorem}
\newtheorem{assumption}{Assumption}
\newtheorem{proposition}{Proposition}

\newtheorem{remark}{Remark}

\newcommand{\blind}{1}

\allowdisplaybreaks

\usetikzlibrary{shapes,decorations,arrows,calc,arrows.meta,fit,positioning}
\tikzset{
    -Latex,auto,node distance =1 cm and 1 cm,semithick,
    state/.style ={ellipse, draw, minimum width = 0.7 cm},
    point/.style = {circle, draw, inner sep=0.04cm,fill,node contents={}},
    bidirected/.style={Latex-Latex,dashed},
    el/.style = {inner sep=2pt, align=left, sloped}
}

\usepackage{amsmath}
\allowdisplaybreaks

\begin{document}

\def\spacingset#1{\renewcommand{\baselinestretch}%
{#1}\small\normalsize} \spacingset{1}


\if1\blind
{

  \title{\Large \bf Identification and efficient estimation of compliance and network causal effects in cluster-randomized trials 
  }
    \author{%
Chao Cheng\thanks{Department of Statistics and Data Science, Washington University in St. Louis, St. Louis, MO. Email: \texttt{chaoc@wustl.edu}.},
Georgia Papadogeorgou\thanks{Department of Statistics, University of Florida, Gainesville, FL. Email: \texttt{gpapadogeorgou@ufl.edu}.},
and Fan Li\thanks{Department of Biostatistics, Yale School of Public Health; Center for Methods in Implementation and Prevention Science, Yale School of Public Health, New Haven, CT.  Email: \texttt{fan.f.li@yale.edu}.}%
}
\date{}

  \maketitle
} \fi

\if0\blind
{
  \bigskip
  \bigskip
  \bigskip
  \begin{center}
    {\Large \bf Identification and efficient estimation of compliance and network causal effects in cluster-randomized trials}
\end{center}
  \medskip
} \fi


\begin{abstract}
Treatment noncompliance is pervasive in infectious disease cluster-randomized trials. Although all individuals within a cluster are assigned the same treatment condition, the treatment uptake status may vary across individuals due to noncompliance.
We propose a semiparametric framework to evaluate the \textit{individual compliance effect} and \textit{network assignment effect} within principal stratum  exhibiting different patterns of noncompliance. The individual compliance effect captures the portion of the treatment effect attributable to changes in treatment receipt, while the network assignment effect reflects the pure impact of treatment assignment and spillover among individuals within the same cluster.
Unlike prior efforts which either empirically identify or interval identify these estimands, we characterize new structural assumptions for nonparametric point identification. We then develop semiparametrically efficient estimators that combine data-adaptive machine learning methods with efficient influence functions to enable more robust inference. Additionally, we introduce sensitivity analysis methods to study the impact under assumption violations, and apply the proposed methods to reanalyze a cluster-randomized trial in Kenya that evaluated the impact of school-based mass deworming on disease transmission.
\end{abstract}

\noindent%

\spacingset{1.75} 


\section{Introduction}

\allowdisplaybreaks

\subsection{Background and research questions}

Cluster-randomized trials (CRTs) are commonly used to evaluate the impact of new interventions administered at the cluster level \citep{hayes2017cluster}. Despite randomization at the cluster level, treatment noncompliance can arise if individuals do not adhere to the allocated treatment for their cluster. In a systematic review of 123 CRTs published in English in 2011, treatment noncompliance was reported in 56 trials \citep{agbla2018reporting}.
Although interventions in CRTs are assigned at the cluster level, noncompliance typically occurs at the individual level, resulting in variations in the individual-level treatment uptake status within clusters. Noncompliance can occur in the context of both health and social science for various reasons. 
For example, when treatments are therapeutic medications, individuals may refuse to take the assigned medication due to concerns about side effects. When treatments are educational programs, individuals may refuse to participate due to lack of motivation.  Moreover, noncompliance is particularly common in the context of \textit{clustered encouragement designs}, a type of CRTs in which individuals in treated clusters are only encouraged (through incentives or invitations) to take the treatment, rather than the treatment itself being  administered  \citep{frangakis2002clustered,forastiere2016identification}. 


In the presence of noncompliance, the intent-to-treat (ITT) analysis serves as a standard approach to study the effectiveness of the cluster-level intervention \citep{wang2024model}, but it only captures the effect of treatment assignment rather than the effect of treatment receipt.
To accommodate noncompliance as an intermediate variable, principal stratification has been introduced to target subgroup estimands conditioning on individuals' compliance types defined by potential intermediate outcomes \citep{frangakis2002principal}. These subgroup estimands, referred to as principal casual effects (PCEs), characterize the causal effects across (i) \textit{compliers}---those who always adhere to their assignment; (ii) \textit{always takers}---those who always receive the treatment regardless of assignment; (iii) \textit{never takers}---those who always receive the control regardless of assignment; and (iv) \textit{defiers}---those who always take the opposite of what assigned. Evaluating the PCE among compliers offers insights into the efficacy of the treatment assignment. 

Although avoiding contamination is often cited as a motivation for adopting a cluster-randomized design, this rationale largely holds only when individual-level noncompliance is minimal or absent. When all individuals within a cluster adhere to their assigned condition, \textit{interference} or \textit{spillover effects} \citep{hudgens2008toward,tchetgen2012causal} are generally not a concern. In the presence of noncompliance, however, interference may occur as individuals differ in their actual treatment receipt. Beyond the PCEs, more subtle estimands concerning interference exist in CRTs. Similar to \cite{forastiere2016identification}, we focus on decomposing the PCE into two components: the \textit{individual compliance effect} (ICE) and the \textit{network assignment effect} (NAE). These components quantify the extent to which the PCE is transmitted through, or independently of, changes in individual treatment uptake, among certain compliance-induced subpopulations. {Here, the ICE reflects the degree to which the treatment achieves its intended impact directly via increasing individual treatment uptake in certain principal strata. In contrast, the NAE captures the part of the treatment effect not explained by an individual’s own treatment uptake. In other words, the NAE integrates both spillover effect arising from the treatment status of other individuals in the same cluster and psychological effects stemming solely from changes in an individual’s assignment status. Hence, the NAE generally cannot be interpreted as a pure, spillover effect.} Nevertheless, evaluating the ICE and NAE offers insights into how treatment assignment operates in conjunction with actual treatment receipt and can help guide the planning and scaling-up of cluster-level interventions.


We contribute to the growing casual inference literature on CRTs with individual-level treatment noncompliance by addressing a critical gap in the identification of complex causal estimands under within-cluster interference. Existing identification methods have relied largely on fully parametric Bayesian mixture models. For example, \citet{frangakis2002clustered} developed Bayesian mixture models for studying PCEs, accounting for within-cluster correlations among observed outcomes and treatment receipt status, but without addressing within-cluster interference. Allowing for within-cluster interference, \citet{forastiere2016identification} described a Bayesian mixture modeling framework with full distributional assumptions to empirically identify the ICE and NAE (alternatively referred to as individual treatment mediated effect and net encouragement effect). However, consistency of their estimators hinges on the correct specification of all parametric models. {Other recent works pursue the exclusion restriction assumption to derive partial identification bounds for spillover effects. Leveraging the network exclusion restriction assumption to rule out direct effects of treatment assignment on outcomes, \citet{park2023assumption} and \citet{kang2019spillover} derived causal bounds for the network effect, yet such bounds may be wide and less informative (a more detailed comparison on the network effects to our ICE and NAE estimands can be found in Remark \ref{remark:compare_to_spillover_efffect}).} 
In this paper, we develop a nonparametric identification strategy for the ICE and NAE estimands based on the principal ignorability assumption---an assumption that was recently formalized for semiparametric principal stratification analysis in the absence of interference \citep{ding2017principal,jiang2022multiply,cheng2026multiply}. This assumption is more likely to hold when sufficient covariate information is available to capture key determinants of treatment uptake and outcomes status.  
Beyond this technical aspect, our framework also aligns with the broader movement toward clarifying target estimands in CRTs. As highlighted in the recent review by \citet{bi2025scoping}, key challenges in causal inference for CRTs include defining the appropriate populations of interest and determining how interference and spillover should be handled. By jointly accommodating informative cluster size \citep{kahan2023estimands} and spillover effects, this work offers a principled foundation for point identifying and interpreting intervention effects in CRTs with real-world patterns of treatment uptake.

The contributions of this work are threefold. First, we show that the ICE and NAE  are nonparametrically identifiable under principal ignorability and monotonicity, thereby establishing point identification for complex estimands that were previously available only under fully parametric models. Second, we develop semiparametrically efficient estimators that combine  machine learning methods to deliver parametric-rate inference \citep{chernozhukov2018double}. To the best of our knowledge, this is the first attempt to study optimal semiparametric estimators in settings where principal ignorability operates in the presence of within-cluster interference. Third, recognizing that principal ignorability is untestable, we contribute a sensitivity approach that quantifies the robustness of conclusions to violations of this identification assumption. {We further develop an R package, \texttt{CRTcompliance}, available in the supplementary material, to implement the proposed methodology.}

\subsection{A motivating data application: the PSDP CRT}

Intestinal helminths infection is a major health concern among children in under-developed regions. The \textit{Primary School Deworming Project} (PSDP) was a CRT in Kenya to evaluate the effects of providing deworming medicine to  child health and education outcomes \citep{miguel2004worms}. The study began in 1998 and included $75$ primary schools, which were randomly assigned to receive the mass deworming treatment in three phases over a three-year period. Our analysis focuses on the initial phase of the PSDP, in which the treatment is defined as assignment to deworming in 1998. In this phase, $K=50$ schools were evenly randomized to conditions. The outcome of interest is a binary indicator of any moderate-to-heavy worm infection measured at the end of the initial phase in 1999 (1 = yes; 0 = no). One-sided noncompliance arose, with approximately 18\% of children in treatment schools not taking the deworming medication. Within-cluster interference is highly likely due to herd immunity: children who did take the medication may have provided protection to peers who did not. Among subgroups of children with different compliance behaviors, we aim to disentangle how much of the treatment effect on infection status operates through changes in individual treatment uptake (i.e., the ICE estimands), and how much is attributable to spillover and pure impact of assignment change (i.e., the NAE estimands). 

\section{Notation, data structure, and causal estimands}

We consider a CRT with $K$ clusters. For cluster $i\in\{1,\dots,K\}$, let $N_i$ be the cluster size, $\bm V_i\in \mathcal V \subseteq \mathbb{R}^{d_V\times 1}$ be a vector of cluster-level baseline covariates, and $A_i\in\{0,1\}$ be a cluster-level treatment with $1$ and $0$ indicating treated and control conditions, respectively. For individual $j\in\{1,\dots,N_i\}$ from cluster $i$, we also define $\bcx_{ij}\in \mathbb{R}^{d_X\times 1}$ as a vector of individual-level baseline covariates, $D_{ij}\in\{0,1\}$ as the treatment uptake status with 1 and 0 indicating the treatment receipt and no treatment, respectively, and $Y_{ij}\in \mathbb{R}$ as the outcome of interest. We let $\bm X_i = [\bm X_{i1}^T,\dots,\bm X_{iN_i}^T]^T$, $\bm Y_i = [Y_{i1},\dots,Y_{iN_i}]^T$ and $\bm D_i = [D_{i1},\dots,D_{iN_i}]^T $ be the collection of individual-level covariates, outcome, treatment uptake status across all individuals in cluster $i$. {Conditioning on $N_i=n_i$, let $\mathcal X(n_i) \subseteq \mathbb{R}^{n_i\times d_X}$, $\mathcal D(n_i)  \subseteq \{0,1\}^{\otimes n_i} $, and $\mathcal Y(n_i) \subseteq \mathbb{R}^{n_i\times 1}$ be the support of $\bm X_i$, $\bm D_i$, and $\bm Y_i$, respectively.} We also define $\bm D_{i (-j)}$ as $\bm D_{i}$ excluding the $j$-th individual, and sometimes equivalently write $\bm D_i=\{D_{ij},\bm D_{i(-j)}\}$. 

For $a\in\{0,1\}$, we let $Y_{ij}(a)$ and $D_{ij}(a)$ be the potential values of outcome and treatment uptake status for individual $j$ in cluster $i$ if this cluster is randomized to condition $a$. Note that we implicitly rule out cross-cluster interference, assuming that an individual’s outcome or treatment receipt does not depend on any information from other clusters. This assumption is standard in CRTs and is reasonable when clusters are not in close geographical proximity. We also denote $Y_{ij}(a,\bm d_{i})$ as the potential outcome of individual $j$ in cluster $i$ if the cluster was randomized to condition $a$ and the treatment uptake status of all individuals in that cluster, $\bm D_{i}$, were set to value $\bm d_i$. Since $\bm D_i=\{D_{ij},\bm D_{i(-j)}\}$, we can re-express $Y_{ij}(a,\bm d_{i})$ as $Y_{ij}(a, d_{ij},\bm d_{i(-j)})$. This notation explicitly state the three causes for individual $(i,j)$'s outcome, including the treatment assignment condition ($a$), this individual's own treatment uptake status ($d_{ij}$), and the remaining cluster members' treatment uptake status ($\bm d_{i(-j)}$).We further assume that $Y_{ij}(a)=Y_{ij}(a, \bm D_{i}(a))=Y_{ij}(a, D_{ij}(a),\bm D_{i(-j)}(a))$ \citep{vanderweele2009conceptual}; i.e., one individual's potential outcome under treatment assignment $a$ is assumed to be equal to this individual's potential outcome if the treatment assignment is set to $a$ and the treatment uptake status for all other cluster members are set to their natural values under treatment assignment $a$.  To proceed, we write the collection of random variables in cluster $i$ as $\mathcal{ U}_i = \{\bcv_i,\bcx_i,  \bm D_i(1),\bm D_i(0),\bm Y_{i}(1,\bm d_i),\bm Y_i(0,\bm d_i)\}$ for all $\bm d_i\in \mathbb{R}^{N_i\times 1}$, where $\bm Y_i(a,\bm d_i) = \{Y_{i1}(a,\bm d_i),\cdots, Y_{iN_i}(a,\bm d_i)\}$. We adopt the following super population framework to characterize the distribution of the full, even though not fully
observed, data vector $\{(N_1,A_1,\mathcal U_1),\dots,(N_K,A_K,\mathcal U_K)\}$:
\begin{assumption}\label{assum:population}
(Super population) The cluster size $N_i$ follows an unknown distribution $\mathbb{P}_{N_i}$ over a finite support $\mathcal N$ on $\mathbb{N}^{+}$. Conditional on $N_i$, 
$\{\mathcal U_i,A_i\}$ follows the joint distribution $\mathbb{P}_{\mathcal U_i,A_i|N_i}=\mathbb{P}_{\mathcal U_i|N_i}\times  \mathbb{P}_{A_i}$ with $\mathbb{P}_{\mathcal U_i|N_i}$ having a finite second moment. 
\end{assumption}

The intent-to-treat (ITT) estimand is often used to assess the overall treatment effectiveness: $\text{ITT}={\E\left[\frac{W_i}{N_i}\sum_{j=1}^{N_i} \left\{Y_{ij}(1)-Y_{ij}(0)\right\}\right]}/{\E[W_i]}$, 
where the expectation is taken over the super population under Assumption \ref{assum:population}. Here, we address the potential for informative cluster size \citep{kahan2023estimands}, and define the ITT estimand as a weighted average treatment effect; that is, $W_i$ is a user-specified cluster-specific weight determining the contribution of each cluster to the ITT estimand, and this weight should be a known function of $N_i$ and at most measured cluster-level covariates $\bm V_i$ \citep{li2025model}. Two typical choices are $W_i=1$ and $W_i=N_i$. The former assigns each cluster equal weight, and leads to the \textit{cluster-average ITT effect}; the latter assigns equal weight to each individual in the study regardless of their cluster membership, and leads to the \textit{individual-average ITT effect}. 

Although the ITT estimand measures the effectiveness, it reflects only the effect of the treatment assignment but not the effect of the actual treatment receipt. Leveraging the principal stratification framework \citep{frangakis2002clustered}, we use the joint potential values of treatment uptake status, $G_{ij}=\{D_{ij}(1),D_{ij}(0)\}\in\{0,1\}^{\otimes 2}$, to divide the population into four subgroups, referred to as principal strata, including always takers ($G_{ij}=\{1,1\}$), compliers ($G_{ij}=\{1,0\}$), never takers ($G_{ij}=\{0,0\}$), and defiers ($G_{ij}=\{0,1\}$). For brevity, we further abbreviate the four principal strata $\{1,1\},\{1,0\},\{0,0\}$, and $\{0,1\}$ as $\texttt{at},\texttt{co},\texttt{nt}$, and $\texttt{de}$, respectively. Let $\mathcal G_{\text{all}} = \{\texttt{at},\texttt{co},\texttt{nt},\texttt{de}\}$ be the set of all principal strata. For a given principal stratum $g\in\mathcal G_{\text{all}}$, we define the \textit{principal causal effect} (PCE)  as 
\begin{equation}\label{eq:pce}
\text{PCE}_{g} = {\E\left[\frac{W_i}{N_i}\displaystyle\sum_{j=1}^{N_i} \mathbb{I}(G_{ij}=g)\left\{Y_{ij}(1)-Y_{ij}(0)\right\}\right]}\Bigg/{\E\left[\frac{W_i}{N_i}\displaystyle\sum_{j=1}^{N_i} \mathbb{I}(G_{ij}=g)\right]},
\end{equation}
where $\mathbb{I}(G_{ij}=g)$ indicates whether individual 
$(i,j)$ belongs to principal stratum $g$. By definition, the numerator of $\text{PCE}_{g}$ calculates the average potential outcome difference between treated and control assignment among individuals in principal stratum $g$, weighted by the pre-specified cluster-specific weight $W_i$. The denominator calculates average number of individuals in principal stratum $g$, weighted by $W_i$ (such that it matches the definition of the ITT estimand). Therefore, $\text{PCE}_{g}$ summarizes the treatment effect for the ``weighted" subgroup of principal stratum $g$. By definition, as long as the choice of weight $W$ is pre-specified, the ITT effect is an average of the PCEs across principal strata such that $\text{ITT} = \displaystyle\sum_{g\in \mathcal G_{\text{all}}} e_{g} \times \text{PCE}_g$, where 
$e_{g}=\E\left[(W_i/N_i)\sum_{j=1}^N \mathbb{I}(G_{ij}=g)\right]\Big/ \mathbb{E}[W_i]$. 

We next decompose $\text{PCE}_g$ into a network assignment effect ($\text{NAE}_g$) and individual compliance effect ($\text{ICE}_g$), defined as
\begin{align*}
\text{NAE}_g & = \frac{\E\left[(W_i/N_i)\displaystyle\sum_{j=1}^{N_i} \mathbb{I}(G_{ij}=g)\left\{Y_{ij}(1,D_{ij}(0),\bm D_{i (-j)}(1))-Y_{ij}(0,D_{i j}(0),\bm D_{i (-j)}(0))\right\}\right]}{\E\left[(W_i/N_i)\displaystyle\sum_{j=1}^{N_i} \mathbb{I}(G_{i j}=g)\right]}, \\
\text{ICE}_g & = \frac{\E\left[(W_i/N_i)\displaystyle\sum_{j=1}^{N_i} \mathbb{I}(G_{i j}=g)\left\{Y_{i j}(1,D_{i j}(1),\bm D_{i (-j)}(1))-Y_{i j}(1,D_{i j}(0),\bm D_{i (-j)}(1))\right\}\right]}{\E\left[(W_i/N_i)\displaystyle\sum_{j=1}^{N_i} \mathbb{I}(G_{i j}=g)\right]}.
\end{align*}
By construction, $\text{NAE}_g$ is defined as the difference in potential outcomes under treated and control cluster assignments but hold the individual's own potential treatment uptake status ($D_{i j}$) fixed at its natural value under the control assignment, averaging over all individuals belonging to principal stratum $g$. Therefore, $\text{NAE}_g$ reflects the effect of changing the cluster assignment condition among individuals in principal stratum $g$, without changing the level of that particular individual treatment. In the context of PSDP study, $\text{NAE}_g$ includes two sources of effects: 
\begin{enumerate*}[label=(\roman*)] 
\item the spillover effect from other same-school children who took the deworming medication,
\item the direct or psychological effect solely due to the change of the school assignment.
\end{enumerate*}
In contrast, $\text{ICE}_g$ compares the difference in potential outcomes, within principal stratum $g$, by changing the individual treatment uptake status from their natural value under control to their natural value under treatment, while fixing the cluster assignment to treatment. In the context of PSDP study, $\text{ICE}_g$ represents the treatment effect operating through each child's treatment uptake status. 

Table \ref{tab:estimands} provides a summary of estimands. Importantly, it should be noted that, for the always takers and never takers, $D_{i j}(1)=D_{i j}(0)$ so that $Y_{i j}(1,D_{i j}(1),\bm D_{i (-j)}(1))=Y_{i j}(1,D_{i j}(0),\bm D_{i (-j)}(1))$, therefore  $\text{ICE}_{\texttt{at}}=\text{ICE}_{\texttt{nt}}=0$. This further implies that $\text{NAE}_{\texttt{at}}=\text{PCE}_{\texttt{at}}$ and $\text{NAE}_{\texttt{nt}}=\text{PCE}_{\texttt{nt}}$. However, $\text{ICE}_{\texttt{co}}$ can differ from 0 among compliers. Additionally, as we will see shortly, defiers are ruled out under the monotonicity assumption.


\begin{table}[h]
\centering
\begin{threeparttable}

\caption{
Summary of the estimands of interest. A ``\checkmark'' indicates that the estimand is substantively meaningful;
a ``0'' indicates that the estimand is zero by construction;
and ``N/A'' indicates that the estimand is not assessable due to structural assumptions.\label{tab:estimands}
}

\begin{tabular}{c|c|c|c|P{7cm}}
\hline
\multicolumn{1}{c|}{Principal Stratum} &
\multicolumn{1}{c|}{$\text{PCE}_g$} &
\multicolumn{1}{c|}{$\text{ICE}_g$} &
\multicolumn{1}{c|}{$\text{NAE}_g$} &
\multicolumn{1}{c}{Remarks} \\
\hline

Compliers 
& \checkmark & \checkmark & \checkmark 
& Both $\text{ICE}_{\mathrm{co}}$ and $\text{NAE}_{\mathrm{co}}$ are well-defined with substantive interest. 
\\ \hline

Never Takers 
& \checkmark & 0 & \checkmark 
& Since $D_{ij}(0)=D_{ij}(1)=0$ holds, we have $\text{ICE}_{\mathrm{ne}}=0$ and $\text{PCE}_{\mathrm{ne}}=\text{NAE}_{\mathrm{ne}}$. 
\\ \hline

Always Takers$^*$ 
& \checkmark & 0 & \checkmark 
& Since $D_{ij}(0)=D_{ij}(1)=1$ holds, we have $\text{ICE}_{\mathrm{at}}=0$ and $\text{PCE}_{\mathrm{at}}=\text{NAE}_{\mathrm{at}}$. 
\\ \hline

Defiers 
& N/A & N/A & N/A 
& Ruled out under monotonicity. 
\\ \hline

\end{tabular}

\vspace{-0.1cm}

\begin{tablenotes}
\footnotesize
\item[$*$] Under strong monotonicity (Assumption \ref{assum:monotonicity}b), the always takers also do not exist, so PCE, ICE, and NAE for the always takers stratum are not analytically assessable in this setting.
\end{tablenotes}

\end{threeparttable}
\end{table}

\vspace{-0.4in}

\section{Assumptions and nonparametric identification}\label{sec:identification}

We consider the following structural assumptions to identify the NAEs and ICEs:
\begin{assumption}\label{assum:consistency}
(Consistency) For any $a\in\{0,1\}$, $\bm d_{i}\in\{0,1\}^{\otimes N_i}$ and $j\in\{1,\cdots, N_i\}$,  we have  $Y_{ij}(a)=Y_{ij}$ and $D_{ij}(a)=D_{ij}$ if $A_i=a$, and $Y_{ij}(a,\bm d_{i})=Y_{ij}$ if $A_i=a$ and $\bcd_{i}=\bm d_{i}$. 
\end{assumption}

\begin{assumption}\label{assum:randomization}
(Cluster randomization) 
$\{A_1,\dots,A_K\}$ are identical and independent draws from a Bernoulli distribution with $\mathbb{P}(A_i=1)=\pi\in (0,1)$.  
\end{assumption}

\begin{assumption}\label{assum:monotonicity}
(Monotonicity) (a) Under standard monotonicity, $D_{ij}(1)\geq D_{ij}(0)$ for all $j\in\{1,\dots,N_i\}$ (b) Under strong monotonicity, $D_{ij}(0)=0$ for all $j\in\{1,\dots,N_i\}$. 
\end{assumption}

\begin{assumption}\label{assum:pi}
(Extended principal ignorability) For all $j\in\{1,\dots,N_i\}$: 
\begin{compactitem}
\item[(a)] $\bbE\left[Y_{i j}(0)\mid G_{i j} = \texttt{co},\bcc_i\right]=\bbE\left[Y_{i j}(0)\mid G_{i j} = \texttt{nt},\bcc_i\right]$;
\item[(b)] $\bbE[Y_{i j}(1)\mid G_{i j} = \texttt{co},\bcc_i]=\bbE[Y_{i j}(1)\mid G_{i j} = \texttt{at},\bcc_i]$;
\item[(c)] $\bbE[Y_{i j}(1,D_{i j}(0),\bcd_{i (-j)}(1))\mid G_{i j} = \texttt{co},\bcc_i]=\bbE[Y_{i j}(1,D_{i j}(0),\bcd_{i (-j)}(1))\mid G_{i j} = \texttt{nt},\bcc_i]$.
\end{compactitem}
\end{assumption}

Assumption \ref{assum:consistency} enables us to connect potential variables to their corresponding observed values, and rules out interference across different clusters. Assumption  \ref{assum:randomization} holds by design, and implies $A_i \perp \{Y_{ij}(a,d_{ij},\bm d_{i(-j)}),\bm D_{i}(1),\bm D_{i}(0),\bm X_i,\bm V_i,N_i\}$. Two types of monotonicity are considered in Assumption  \ref{assum:monotonicity}. The standard monotonicity rules out existence of  defiers \citep{frangakis2002principal}. In the one-sided noncompliance scenario (e.g., the PSDP CRT), individuals in control clusters have no access to the treatment. In this case, strong monotonicity is satisfied, and rules out both defiers and always takers. Monotonicity allows us to identify the principal score \citep{ding2017principal}, $e_{ij}(g,\bm X_i,\bm V_i, N_i) = \mathbb{P}(G_{ij}=g\mid \bm X_i,\bm V_i, N_i)$---the probability of individual stratum membership given observed baseline information in cluster $i$. For simplicity, we write $\bm C_i = \{\bm X_i,\bm V_i, N_i\}$ and $e_{ij}(g,\bm X_i,\bm V_i, N_i)\equiv e_{ij}(g,\bm C_i)$. Following \cite{cheng2025identification}, we have the unified identification formula of $e_{ij}(g,\bm C_i)$ under both monotonicity assumptions. Specifically, defining  $p_{ij}(a,d,\bm C_{i}) = \mathbb{P}(D_{ij}=d\mid A_{i}=a,\bm C_i)$, 
we have
\begin{equation}\label{eq:principal_score}
e_{ij}(g,\bm C_i) = p_{ij}(a^{\dagger},d^{\dagger},\bm C_{i})-h\times p_{ij}(0,1,\bm C_i),
\end{equation}
where $\{a^\dagger,d^\dagger,h\}=[\{0,1,0\},\{1,1,1\},\{1,0,0\},\{0,1,1\}]$ if $g=[\texttt{at},\texttt{co},\texttt{nt},\texttt{de}]$ respectively. 

{Assumption \ref{assum:pi} extends the principal ignorability assumption developed for the independent data setting \citep{ding2017principal,lu2026principal} to CRTs. Specifically, conditional on $\bcc_i$, Assumption \ref{assum:pi}(a) requires that the mean potential outcome is mean exchangeable across compliers and never-takers, and Assumption 5(b) requires that the treated potential outcome is mean exchangeable across compliers and always-takers. Assumption \ref{assum:pi}(a)--(b) are sufficient to identify the PCEs across all principal strata but not $\text{ICE}_{\texttt{co}}$ and $\text{NAE}_{\texttt{co}}$. For these two finer estimands, Assumption \ref{assum:pi}(c) is additionally invoked to ensure that $Y_{i j}(1,D_{i j}(0),\bcd_{i (-j)}(1))$ is mean exchangable across compliers and never-takers given baseline covariates. We note that a stronger, possibly more interpretable, version of Assumption \ref{assum:pi} is
$
Y_{ij}(a,D_{ij}(a^*),\bm D_{i(-j)}(a))\perp G_{ij}\mid\{\bm X_i,\bm V_i,N_i\},
$ for any  $a,a^*\in\{0,1\}$. This stronger assumption states that the principal stratum membership behaves as if conditionally randomized with respect to the relevant nested potential outcomes given measured covariates.

In Web Appendix A.1, we have provided a structural model to illustrate the restrictions of Assumption \ref{assum:pi} on the data-generating process. In the context of the PSDP CRT, Assumption \ref{assum:pi} is plausible if: (1) sufficient information is collected on baseline characteristics that influence both student’s noncompliance status and health outcomes, and no unmeasured confounding exists after conditioning on the collected baseline information; (2) there are no post-baseline covariates that jointly influence student's noncompliance status and health outcomes; (3) after conditioning on the collected baseline covariates, there are no residual common factors that induce dependence in compliance behavior across students within the same school. Condition (3) requires that the observed baseline covariates adequately capture the major determinants of student's treatment receipt status, so that one student's compliance status does not carry residual information about other same-cluster children's compliance status beyond that already explained by the observed covariates. Assumption \ref{assum:pi} is untestable; therefore, in Section \ref{sec:sensitivity}, we develop a sensitivity approach to examine the robustness of findings to its violations.}

Finally, Assumption \ref{assum:positivity} states that the relevant principal strata are non-trivial, {and included a technical condition that $W_i$ is bounded away from infinity.}

\begin{assumption}\label{assum:positivity}
{(Positivity and boundedness) 
 (a) For any $n_i\in \mathcal N$, $\bm v_i \in \mathcal V$, $\bm x_i \in \mathcal X(n_i)$, $1\leq j\leq n_i$, there exist $0<b_1,b_2<0.5$ such that $b_1< e_{ij}(\texttt{co},\bm c_i)<1-b_1$ and $b_1< e_{ij}(\texttt{nt},\bm c_i)<1-b_1$. Under standard monotonicity, additionally assume $b_1< e_{ij}(\texttt{at},\bm c_i)<1-b_1$. (b) There exist $0<b_3<b_4<\infty$ such that the weight satisfies $b_3<W_i<b_4$.}
\end{assumption}

For a given $g\in\{\texttt{at},\texttt{co},\texttt{nt}\}$ under standard monotonicity, or a given $g\in\{\texttt{co},\texttt{nt}\}$ under strong monotonicity, it is sufficient to identify
$$
\theta_g(a,a^*) = \E\left[\frac{W_i}{N_i}\displaystyle\sum_{j=1}^{N_i} \mathbb{I}(G_{i j}=g)Y_{i j}(a,D_{ij}(a^*),\bm D_{i(-j)}(a))\right]\Bigg/{\E\left[\frac{W_i}{N_i}\displaystyle\sum_{j=1}^{N_i} \mathbb{I}(G_{i j}=g)\right]}
$$
for any $\{a,a^*\}\in [\{1,1\},\{1,0\},\{0,0\}]$ if $g=\texttt{co}$, $\{a,a^*\}\in [\{1,0\},\{0,0\}]$ if $g=\texttt{at}$, and $\{a,a^*\}\in [\{1,0\},\{0,0\}]$ if $g=\texttt{nt}$.
Based on $\theta_g(a,a^*)$, we can obtain all NAEs and ICEs in Table \ref{tab:estimands} by $\text{NAE}_{\texttt{co}} = \theta_{\texttt{co}}(1,0)-\theta_{\texttt{co}}(0,0)$, $\text{NAE}_{\texttt{nt}} = \theta_{\texttt{nt}}(1,0)-\theta_{\texttt{nt}}(0,0)$, 
$\text{NAE}_{\texttt{at}} = \theta_{\texttt{at}}(1,0)-\theta_{\texttt{at}}(0,0)$, $\text{ICE}_{\texttt{co}} = \theta_{\texttt{co}}(1,1)-\theta_{\texttt{co}}(1,0)$,
whereas $\text{ICE}_{\texttt{at}}=\text{ICE}_{\texttt{nt}}=0$ by construction. Theorem \ref{thm:identification} provides the nonparametric identification formula.

\begin{theorem}\label{thm:identification}
Under Assumptions \ref{assum:population}--\ref{assum:positivity}, 
$\theta_g(a,a^*)$ is nonparametrically identified as
$$
\theta_g(a,a^*) = {\E\left[\frac{W_i}{N_i}\displaystyle\sum_{j=1}^{N_i} e_{i j}(g,\bm C_i) \times \mu_{i j}(a,d^*,\bm C_i)\right]}\Big/{\E\left[\frac{W_i}{N_i}\displaystyle\sum_{j=1}^{N_i} e_{i j}(g,\bm C_i)\right]},
$$
where $\mu_{i j}(a,d,\bm C_i)=\E[Y_{i j}\mid A_i=a,D_{i j}=d,\bm C_i]$ and  $e_{i j}(g,\bm C_i)$ is identified in \eqref{eq:principal_score}. Here, $d^*=[1,a^*,0]$ if $g=[\texttt{at},\texttt{co},\texttt{nt}]$, respectively. 
\end{theorem}

{\begin{remark}[Connection to network effects]
\label{remark:compare_to_spillover_efffect}
\cite{park2023assumption} and \cite{kang2019spillover} studied network effects in CRTs under network exclusion restriction, which rules out effect of assignment on the outcome except through changes in treatment receipt; that is 
$Y_{ij}(1,D_{ij}(a),\bm D_{i(-j)}(a))
=
Y_{ij}(0,D_{ij}(a), \bm D_{i(-j)}(a))
=
Y_{ij}\langle D_{ij}(a),\bm D_{i(-j)}(a)\rangle$ for $a\in\{0,1\}$. In our setup, the network effect estimand can be written as
$$
\mathrm{NetEff}_g
=
\frac{
\E\left[
W_i/N_i\sum_{j=1}^{N_i}
\mathbb{I}(G_{ij}=g)
\left\{
Y_{ij}\langle D_{ij}(1),\bm D_{i(-j)}(1)\rangle
-
Y_{ij}\langle D_{ij}(0),\bm D_{i(-j)}(0)\rangle
\right\}
\right]
}{
\E\left[
W_i/N_i\sum_{j=1}^{N_i}
\mathbb{I}(G_{ij}=g)
\right]
},
$$
for $g\in\{\texttt{at},\texttt{co},\texttt{nt}\}$. Because the network exclusion restriction implies
$Y_{ij}\langle D_{ij}(a),\bm D_{i(-j)}(a)\rangle
=
Y_{ij}(a,D_{ij}(a),\bm D_{i(-j)}(a))
=
Y_{ij}(a)$,
we have $\mathrm{NetEff}_g=\mathrm{PCE}_g$. Thus, under the network exclusion restriction, $\mathrm{PCE}_g$ is itself interpreted as a network effect estimand.

Under the network exclusion restriction,  monotonicity (Assumption \ref{assum:monotonicity}), and a bounded monotone outcome assumption ($0\leq Y_{ij}\langle 0,\bm d_{i(-j)} \rangle \leq Y_{ij}\langle 1,\bm d_{i(-j)} \rangle\leq 1$),
\cite{park2023assumption} developed partial-identification bounds for the network effects. If network exclusion restriction is violated, these bounds can be viewed as bounds for $\mathrm{PCE}_g$. In contrast, by imposing extended principal ignorability (Assumption~\ref{assum:pi}), we can obtain nonparametric point identification of $\mathrm{PCE}_g$ and further decomposes the complier PCE into finer estimands: 
$\mathrm{PCE}_{\texttt{co}}=\mathrm{ICE}_{\texttt{co}}+\mathrm{NAE}_{\texttt{co}}$.
For always-takers and never-takers, $\mathrm{ICE}_{\texttt{at}}=\mathrm{ICE}_{\texttt{nt}}=0$ by definition, so that $\mathrm{PCE}_{\texttt{at}}=\mathrm{NAE}_{\texttt{at}}$ and $\mathrm{PCE}_{\texttt{nt}}=\mathrm{NAE}_{\texttt{nt}}$.
\end{remark}}

\vspace{-0.2in}

{\begin{remark}[Connection to causal mediation]\label{rem:mediation}
Although we use $D_{ij}$ to define principal strata, a complementary perspective is to view $D_i$ as a mediator. This creates a connection to causal mediation analysis for CRTs with interference \citep{vanderweele2013mediation,cheng2025identification}, where the ITT estimand is decomposed into the natural direct effect (NDE), individual mediation effect (IME), and spillover mediation effect (SME). As we show in Web Appendix A.2, the IME is a weighted average of the individual compliance effects across strata: $\mathrm{IME}=\sum_{g\in\mathcal G_{\mathrm{all}}} e_g \times \mathrm{ICE}_g$. Similarly, the sum of the NDE and SME is a weighted average of the network assignment effects, with $\mathrm{NDE}+\mathrm{SME}=\sum_{g\in\mathcal G_{\mathrm{all}}} e_g \times \mathrm{NAE}_g$. Thus, our estimands are finer components of causal mediation estimands in CRTs.
\end{remark}}

\vspace{-0.2in}

\section{Doubly robust and efficient estimation}\label{sec:estimation}
\subsection{Moment estimators}\label{sec:estimation_mo}

The identification formula in Theorem \ref{thm:identification} suggests the following moment estimator:
$$
\widehat{\theta}_g^{\text{mo}}(a,a^*) = \left\{\frac{1}{K}\sum_{i=1}^K\frac{W_i}{N_i}\displaystyle\sum_{j=1}^{N_i} \widehat e_{ij}(g,\bm C_i) \widehat{\mu}_{ij}(a,d^*,\bm C_i)\right\}\Big/\left\{\frac{1}{K}\sum_{i=1}^K\frac{W_i}{N_i}\displaystyle\sum_{j=1}^{N_i} \widehat e_{ij}(g,\bm C_i)\right\},
$$
which replaces the expectation operator ``$\E$"  in the identification formula with the empirical cluster-average operator ``$\frac{1}{K}\sum_{i=1}^K$" and also replaces $\{{\mu}_{ij}(a,d^*,\bm C_i),e_{ij}(g,\bm C_i)\}$ with their corresponding estimators. As $e_{ij}(g,\bm C_i)$ is function of $p_{ij}(a,d,\bm C_i)$ in \eqref{eq:principal_score}, we need to estimate the following two nuisance functions, $h_{\text{nuisance}} = \{p_{ij}(a,d,\bm C_i),\mu_{ij}(a,d,\bm C_i)\}$ for any $a,d\in\{0,1\}$, in order to calculate $\widehat{\theta}_g^{\text{mo}}(a,a^*)$. Familiar parametric models can be used to estimate $h_{\text{nuisance}}$. {For $p_{ij}(a,d,\bm C_i) := \mathbb{P}(D_{ij}=d\mid A_i=a,\bm C_i)$, one can specify a logistic regression of $D_{ij}$ on $A_i$ and $\bm C_i=\{\bm X_i, \bm V_i, N_i\}$, where the regression coefficients can be obtained by generalized estimating equations with a chosen working correlation structure. Because the dimension of $\bm X_i$ varies with the cluster size, a summary function can be used to map $\bm X_i$ into a fixed-dimensional covariate vector $\bm S_{ij}$ for regression adjustment \citep{ogburn2024causal,cheng2024semiparametric}. For example, one may specify $\bm S_{ij}=\bm X_{ij}\in\mathbb{R}^{d_X\times 1}$ without covariate interference, or 
$
\bm S_{ij}
=
\left\{
\bm X_{ij},
\frac{1}{N_i-1}\sum_{l\neq j}^{N_i}\bm X_{il}
\right\}
\in\mathbb{R}^{2d_X\times 1},
$
which allows other covariate interference on $D_{ij}$ through a cluster-level average.} Analogously, one can specify a linear or generalized linear regression for $\mu_{ij}(a,d,\bm C_i) = \E[Y_{ij}\mid A_i=a,D_{ij}=d,\bm C_i]$, and the coefficients can be similarly obtained by generalized estimating equations with a suitable working correlation structure.

\subsection{Doubly robust estimators}\label{sec:estimation_dr}

The moment estimator $\widehat{\theta}_g^{\text{mo}}(a,a^*)$ carries asymptotic bias when either the parametric models for the principal score or the outcome mean is misspecified. We therefore develop a doubly robust estimator to provide protection against partial model misspecifications. This development leverages the efficient influence function (EIF) of $\theta_g(a,a^*)$, which is a mean-zero function capturing the key features of the data-generating process relevant for estimating this parameter. The EIF also quantifies the semiparametric efficiency bound for $\theta_g(a,a^*)$ and serves as the foundational tool for developing estimators that can achieve this bound \citep{bickel1993efficient}. To proceed, let $\bm O_i = \{N_i, \bm V_i, \bm X_i, A_i, \bm D_i, \bm Y_i\}$ be the observed data in cluster $i$, and define $\mathcal M_{\text{np}}$ as the nonparametric model of $\bm O_i$, which does not place any restrictions on the likelihood of $\bm O_i$ except for the known treatment assignment probability. Under a nonparametric model, we first derive the EIF of $\theta_g(a,a^*)$ under $\mathcal M_{\text{np}}$ as follows.

\begin{theorem}\label{thm:eif}
Under Assumptions \ref{assum:population}--\ref{assum:positivity}, 
the EIF of $\theta_g(a,a^*)$ under $\mathcal M_{\text{np}}$ is given by
$$
\mathcal D_{g}(a,a^*;\bm O_i) = \frac{({W_i}/{N_i})\sum_{j=1}^{N_i}\left\{\psi_{i j}^{(1)}(a,d^*,a^{\dagger},d^{\dagger},h;\bm O_i)-\psi_{i j}^{(2)}(a^{\dagger},d^{\dagger},h;\bm O_i)\times \theta_g(a,a^*)\right\}}{\E[(W_i/N_i)\sum_{j=1}^{N_i} e_{i j}(g,\bm C_i)]},
$$
where 
\begin{align*}
\psi_{i j}^{(1)}(a,d^*,a^{\dagger},d^{\dagger},h;\bm O_i) = &  \frac{\bbI(A_i=a,D_{ij}=d^*)}{\pi_a \times p_{i j}(a,d^*,\bcc_i)}\{Y_{i j}-\mu_{i j}(a,d^*,\bcc_i)\}\{p_{i j}(a^\dagger,d^\dagger,\bcc_i) - h p_{i j}(0,1,\bcc_i)\} \\
&  + \psi_{i j}^{(2)}(a^{\dagger},d^{\dagger},h,\bm O_i) \mu_{i j}(a,d^*,\bcc_i), \\
\psi_{i j}^{(2)}(a^{\dagger},d^{\dagger},h,\bm O_i) =   &  \frac{\bbI(A_i=a^\dagger)\{\bbI(D_{i j}=d^\dagger)-p_{i j}(a^\dagger,d^\dagger,\bcc_i)\}}{\pi_{a^\dagger}}-h\frac{(1-A_i)\{D_{i j}-p_{i j}(0,1,\bcc_i)\}}{\pi_{0}} \\
& + p_{i j}(a^\dagger,d^\dagger,\bcc_i) - h p_{i j}(0,1,\bcc_i).
\end{align*}
Here, $\pi_a = \mathbb{P}(A_i=a)=\pi^a \times (1-\pi)^{1-a}$, $\{d^*, a^\dagger,d^\dagger,h\}=[\{1,0,1,0\},\{a^*,1,1,1\},\{0,1,0,0\}]$ if $g=[\texttt{at},\texttt{co},\texttt{nt}]$, respectively. Moreover, the semiparametric efficiency bound for the estimation of $\theta_g(a,a^*)$ is $\E[\{\mathcal D_{g}(a,a^*;\bm O_i)\}^2]$. 
\end{theorem}
Theorem \ref{thm:eif} suggests that the EIF is a function of $h_{\text{nuisance}}$ with a constant denominator $\E[(W_i/N_i)\sum_{j=1}^{N_i} e_{i j}(g,\bm C_i)]$. Since the EIF has mean zero, we can solve the moment condition $\E[\mathcal D_{g}(a,a^*;\bm O_i)]=0$ with respect to $\theta_g(a,a^*)$, and obtain
$$
\theta_g(a,a^*) = \E\left[\frac{W_i}{N_i}\sum_{j=1}^{N_i}\psi_{i j}^{(1)}(a,d^*,a^{\dagger},d^{\dagger},h;\bm O_i)\right]\Big/\E\left[\frac{W_i}{N_i}\sum_{j=1}^{N_i}\psi_{i j}^{(2)}(a^{\dagger},d^{\dagger},h;\bm O_i)\right].
$$
This population-level formula motivates the following plug-in estimator:
\begin{equation}\label{eq:doubly_robust_estimator}
\widehat\theta_g^{\text{dr}}(a,a^*) = \frac{1}{K}\sum_{i=1}^K\frac{W_i}{N_i}\sum_{j=1}^{N_i}\widehat \psi_{i j}^{(1)}(a,d^*,a^{\dagger},d^{\dagger},h;\bm O_i)\Big/\frac{1}{K}\sum_{i=1}^K\frac{W_i}{N_i}\sum_{j=1}^{N_i}\widehat \psi_{ij}^{(2)}(a^{\dagger},d^{\dagger},h;\bm O_i),
\end{equation}
where $\widehat \psi_{i j}^{(1)}(a,d^*,a^{\dagger},d^{\dagger},h;\bm O_i)$ and $\widehat \psi_{i j}^{(2)}(a^{\dagger},d^{\dagger},h;\bm O_i)$ are calculated by replacing the unknown nuisance functions $h_{\text{nuisance}}$ in $ \psi_{i j}^{(1)}(a,d^*,a^{\dagger},d^{\dagger},h;\bm O_i)$ and $ \psi_{i j}^{(2)}(a^{\dagger},d^{\dagger},h;\bm O_i)$ based on their parametric model estimates described in Section \ref{sec:estimation_mo}. 

To study the properties of $\widehat\theta_g^{\text{dr}}(a,a^*)$, we use $\mathcal M_e$ for the parametric submodel of $\mathcal M_{np}$ with correct specification of $\widehat p_{ij}(a,d,\bm C_i)$ and other likelihood components unspecified. Similarity, we use $\mathcal M_o$ for the parametric submodel of $\mathcal M_{np}$ with correct specification of $\widehat \mu_{ij}(a,d,\bm C_i)$. We use ``$\cup$" and ``$\cap$" to denote union and intersection of submodels. 
Theorem \ref{thm:dr} summarizes the asymptotic behavior of $\widehat\theta_g^{\text{dr}}(a,a^*)$.
\begin{theorem}\label{thm:dr}
Under Assumptions \ref{assum:population}--\ref{assum:positivity} and regularity conditions in Web Appendix A.3, if $\mathcal M_e \cup \mathcal M_o$ is correctly specified, then $\widehat\theta_g^{\text{dr}}(a,a^*)$ is $\sqrt{K}$-consistent and asymptotically normal in the sense that  $\sqrt{K}\left\{\widehat\theta_g^{\text{dr}}(a,a^*)-\theta_g(a,a^*)\right\}$ converges to a zero-mean normal distribution with finte variance $V_{dr}$, with the explicit form of $V_{dr}$ defined in {Web Appendix A.3}. Furthermore, if $\mathcal M_e \cap \mathcal M_o$ is correctly specified, $V_{dr} = \E[\{\mathcal D_{g}(a,a^*;\bm O)\}^2]$ equals the semiparametric efficiency lower bound defined by the variance of the efficient influence function.  
\end{theorem}
By Theorem \ref{thm:dr}, $\widehat\theta_g^{\text{dr}}(a,a^*)$ offers double protection against model misspecification in the sense that it is consistent and asymptotically normal if at least one of the principal score model for $p_{ij}(a,d,\bm C_i)$ and the outcome model for $\mu_{ij}(a,d,\bm C_i)$ is correctly specified. We therefore refer to $\widehat\theta_g^{\text{dr}}(a,a^*)$ as the doubly robust estimator.  Moreover, if both parametric models are correct, $\widehat\theta_g^{\text{dr}}(a,a^*)$ is semiparametrically efficient and attains the variance lower bound among a class of regular and asymptotically linear estimators for the target estimand. Finally, the asymptotic variance $V_{dr}$ in Web Appendix A.3 has a rather cumbersome expression. Therefore, we consider a model-agnostic nonparametric cluster bootstrap to construct confidence intervals \citep{field2007bootstrapping}. First, we resample the $K$ clusters with replacement, and re-calculate $\widehat\theta_g^{\text{dr}}(a,a^*)$ based on the resampled data. We then repeat the previous process for a large number of replications to construct a bootstrap distribution of $\widehat\theta_g^{\text{dr}}(a,a^*)$. The \( \alpha/2 \) and \( (1 - \alpha/2) \) quantiles of the bootstrap distribution are used as the \( (1 - \alpha) \times 100\% \) confidence interval.

\subsection{Nonparametric efficient estimator via machine learning}\label{sec:np}

To avoid potential model misspecification bias, we further leverage data-adaptive machine learners to estimate $h_{\text{nuisance}}$ when implementing the doubly robust estimator. \cite{phillips2023practical} listed nonparametric models and machine learning algorithms available in the \texttt{SuperLearner} R package, which can be used to estimate $p_{ij}(a,d,\bm C_i)$ and $\mu_{ij}(a,d,\bm C_i)$. Although such data-adaptive methods are generally less vulnerable to model misspecification than parametric methods, they can be more susceptible to overfitting, presenting challenges for inference at parametric rates without modifications. We adopt the cross-fitting procedure below to circumvent the overfitting bias and obtain the debiased machine learning estimates \citep{chernozhukov2018double}. To proceed, we evenly and randomly split the cluster index set $\{1,\cdots, K\}$ into $L$ non-overlapping groups (e.g., $L=5$). For each $l\in\{1,\dots,L\}$, we denote $\mathcal N_l$ as the validation sample including indexes in the $l$-th group, and denote $\mathcal N_{-l}=\{1,\dots,K\}\backslash \mathcal N_l $ as the training sample including the indexes in other groups. 
Then, for each cluster $i$ in $\mathcal N_l$, we calculate $\widehat \psi_{i j}^{(1)}(a,d^*,a^{\dagger},d^{\dagger},h;\bm O_i)$ and $\widehat \psi_{i j}^{(2)}(a^{\dagger},d^{\dagger},h;\bm O_i)$ for all $j$ in cluster $i$ with nuisance functions $h_{nuisance}$ estimated from machine learners based on clusters in the training sample $\mathcal N_{-l}$, denoted by $\widehat h_{nuisance,l}$. We then repeat the previous step from $\mathcal N_1$ to $\mathcal N_L$ to calculate $\widehat \psi_{i j}^{(1)}$ and $\widehat \psi_{i j}^{(2)}$ among all individuals in all clusters. Finally, we substitute $\widehat \psi_{i j}^{(1)}$ and  $\widehat \psi_{i j}^{(2)}$ into \eqref{eq:doubly_robust_estimator} to obtain the nonparametric efficient, debiased machine learning estimator $\widehat \theta_{g}^{\text{np}}(a,a^*)$. {Define $\|\cdot\|$ as the $L_2(\mathbb{P})$-norm such that $\|m(\bm O)\| = \{\E[m(\bm O)^2]\}^{1 / 2}$ for a real-valued function $m(\cdot)$. We describe the asymptotic behavior of $\widehat \theta_{g}^{\text{np}}(a,a^*)$ below.}


{
\begin{theorem}\label{thm:np}
Suppose Assumptions \ref{assum:population}--\ref{assum:positivity} hold. Under standard monotonicity, the following two conditions hold for all $\{a,d\}\in\{0,1\}^{\otimes 2}$ and $\{a',d'\}\in\{0,1\}^{\otimes 2}$. Under strong monotonicity, they hold for $\{a,d\}\in[\{1,1\},\{1,0\}]$ and $\{a',d'\}\in[\{1,1\},\{1,0\},\{0,0\}]$.
\begin{compactitem}
\item[(a)] (Boundedness) For any $n_i\in \mathcal N$, $\bm v_i \in \mathcal V$, $\bm x_i \in \mathcal X(n_i)$, $j\in\{1,\dots,n_i\}$, and $l\in\{1,\dots,L\}$, there exist $0<b_1<0.5$ and $0<b_2<\infty$ such that $b_1<\widehat p_{ij,l}(a,d,\bm c_i)<1-b_1$, $|\mu_{ij}(a',d',\bm c_i)|<b_2$, and $|\widehat \mu_{ij,l}(a',d',\bm c_i)|<b_2$.
\item[(b)] (Convergence rate) Let $r_{p,K}=O(1)$ and $r_{\mu,K}=O(1)$ be two deterministic sequences depending on $K$. For any $l\in\{1,\dots,L\}$, suppose that  $\left\|\max_{1\leq j \leq N_i}\left|(\widehat p_{ij,l}-p_{ij})(a,d,\bm C_i)
\right|\right\|
=
O_{\mathbb P}(r_{p,K}) $ and $\left\|\max_{1\leq j \leq N_i} \left|(\widehat \mu_{ij,l}-\mu_{ij})(a',d',\bm C_i)\right|\right\| = O_\Prob(r_{\mu,K})$.
\end{compactitem}
Then, $\widehat \theta_{g}^{\text{np}}(a,a^*)$ is consistent if either $r_{p,K}=o(1)$ or $r_{\mu,K} = o(1)$. Furthermore, if both $r_{p,K}=r_{\mu,K} = o(1)$ with $r_{p,K} r_{\mu,K} = o(K^{-1/2})$, then $\sqrt{K}(\widehat \theta_{g}^{\text{np}}(a,a^*)- \theta_{g}(a,a^*))$ converges to a zero-mean normal distribution with finite variance $\E[\{\mathcal D_g(a,a^*;\bm O_i)\}^2]$. 
\end{theorem}}

{Theorem~\ref{thm:np} invokes two regularity conditions on the nuisance functions and their estimators. Condition~(a) states boundedness assumption of the nuisance functions and their corresponding estimates. This condition prevents the propensity score weights, $1/\widehat p_{ij,l}(a,d^*,\bm C_i)$, from becoming exceesively large and unstable. Condition~(b) specifies the nuisance convergence rates under the $L_2(\mathbb P)$ norm. Specifically, $r_{p,K}$ and $r_{\mu,K}$ denote the convergence rates of the maximum absolute estimation errors within each cluster for $\widehat p_{ij}(a,d,\bm C_i)$ and $\widehat\mu_{ij}(a,d,\bm C_i)$. Thus, $r_{p,K}=o(1)$ corresponds to consistency of the principal score estimator in the $L_2(\mathbb{P})$-norm, and $r_{\mu,K}=o(1)$ corresponds to consistency of the outcome regression estimator in the $L_2(\mathbb{P})$-norm. Consequently, consistency of $\widehat\theta_g^{\mathrm{np}}(a,a^*)$ follows if either nuisance component is consistently estimated, while the stronger product rate condition $r_{p,K}r_{\mu,K}=o(K^{-1/2})$ is required for asymptotic normality and efficiency. In other words, Theorem \ref{thm:np} suggests that a $o_p(K^{-1/4})$-type convergence rate among both nuisance functions are sufficient to ensure that $\widehat \theta_{g}^{\text{np}}(a,a^*)$ is $\sqrt{K}$-consistent and asymptotically normal, with its asymptotic variance achieving the semiparametric efficiency lower bound.} Finally, since the bootstrapping is not theoretically justified with machine learning-based estimators (unless stronger assumptions are considered), we use the empirical variance of the EIF for consistent variance calculation:
\begin{equation*}
\widehat{\text{Var}}(\widehat \theta_{g}^{\text{np}}(a,a^*)) \!=\! \frac{\frac{1}{K^2}\sum_{i=1}^K\left\{\frac{W_i}{N_i}\sum_{j=1}^{N_i} \Big[\widehat \psi_{ij}^{(1)}(a,d^*,a^{\dagger},d^{\dagger},h;\bm O_i)\!-\!\widehat \psi_{i j}^{(2)}(a^{\dagger},d^{\dagger},h;\bm O_i)\!\times\! \widehat \theta_{g}^{\text{np}}(a,a^*)\Big]\right\}^2}{\left\{\frac{1}{K}\sum_{i=1}^K\frac{W_i}{N_i}\sum_{j=1}^{N_i}\widehat \psi_{i j}^{(2)}(a^{\dagger},d^{\dagger},h;\bm O_i)\right\}^2},
\end{equation*}
where $\widehat \psi_{ij}^{(1)}(a,d^*,a^{\dagger},d^{\dagger},h;\bm O_i)$ and $\widehat \psi_{i j}^{(2)}(a^{\dagger},d^{\dagger},h;\bm O_i)$ are recycled from those in constructing $\widehat \theta_{g}^{\text{np}}(a,a^*)$. 
A Wald-type confidence interval can then be constructed based on $\widehat{\text{Var}}(\widehat \theta_{g}^{\text{np}}(a,a^*))$.

{As a technical note, Web Appendix A.4 provides a more detailed investigation of the estimation error of $\widehat{\theta}_g^{\mathrm{np}}(a,a^*)$, by showing that its convergence rate is governed by the number of clusters $K$, cluster size $N_i$, and within-cluster correlation of the individual-level EIF score, $\xi_{ij}(a,a^*) = \psi_{i j}^{(1)}(a,d^*,a^{\dagger},d^{\dagger},h;\bm O_i)-\psi_{i j}^{(2)}(a^{\dagger},d^{\dagger},h;\bm O_i)\times \theta_g(a,a^*)$. Interestingly, when $\xi_{ij}(a,a^*)$ are only weakly correlated, we have $\widehat{\theta}_g^{\mathrm{np}}(a,a^*) - \theta_g(a,a^*) \approx O_p\{\sqrt{K^{-1}\E(1/N_i)}\}$. Thus, the proposed estimator can still exhibit stable finite-sample performance in CRTs with a moderate number of clusters but relatively large cluster sizes.}

\subsection{Estimation of the ICEs and NAEs}

After obtaining estimates of $\theta_g(a,a^*)$, ICEs and NAEs can be straightforwardly estimated, following the discussion in Section \ref{sec:identification}. 
In Web Appendix A.5, we further summarize the asymptotic behavior of the doubly robust estimator and nonparametric efficient estimator for the NAEs and ICEs. {In addition, although our estimands are defined on the mean difference scale, it is straightforward to formulate our estimands on summary measure scales. For binary outcomes, one may instead work on the risk ratio scale and define $\text{NAE}_{\texttt{co}}^{\text{RR}} = \theta_{\texttt{co}}(1,0)\Big/\theta_{\texttt{co}}(0,0)$, $\text{NAE}_{\texttt{nt}}^{\text{RR}} = \theta_{\texttt{nt}}(1,0)\Big/\theta_{\texttt{nt}}(0,0)$, $\text{NAE}_{\texttt{at}}^{\text{RR}} = \theta_{\texttt{at}}(1,0)\Big/\theta_{\texttt{at}}(0,0)$ and $\text{ICE}_{\texttt{co}}^{\text{RR}} = \theta_{\texttt{co}}(1,1)\Big/\theta_{\texttt{co}}(1,0)$. Estimation of these ratio scale causal effects follows directly from the estimators of $\theta_g(a,a^*)$, and is therefore omitted for brevity.}

\section{Sensitivity analysis under assumption violation}
\label{sec:sensitivity}


Because the extended principal ignorability assumption (Assumption \ref{assum:pi}) is empirically unverifiable from the observed data alone, we develop a sensitivity analysis strategy to evaluate the robustness under its departure. For simplicity, we assume the conditional means of potential outcome to be positive; 
if the outcome is negative, transformations can be made to satisfy this condition. We propose the following set of sensitivity function: 
\begin{align*}
\alpha_{i j}(\bcc_i) := & {\bbE\left[Y_{i j}(0)\mid G_{i j} = \texttt{co},\bcc_i\right]}\Big/{\bbE\left[Y_{i j}(0)\mid G_{i j} = \texttt{nt},\bcc_i\right]}, \\
\beta_{i j}(\bcc_i) := & {\bbE[Y_{i j}(1)\mid G_{i j} = \texttt{co},\bcc_i]}\Big/{\bbE[Y_{i j}(1)\mid G_{i j} = \texttt{at},\bcc_i]}, \\
\gamma_{i j}(\bcc_i) := & {\bbE[Y_{i j}(1,D_{i j}(0),\bcd_{i (-j)}(1))\mid G_{i j} = \texttt{co},\bcc_i]}\Big/{\bbE[Y_{i j}(1,D_{i j}(0),\bcd_{i (-j)}(1))\mid G_{i j} = \texttt{nt},\bcc_i]}.
\end{align*}
Our sensitivity analysis framework is based on allowing for deviations of these sensitivity functions from 1. Intuitively,  $\alpha_{i j}(\bcc_i)$ represents the ratio of the control potential outcome between compliers and never takers, given covariates level $\bm C_i$. In the PSDP study, $\alpha_{i j}(\bcc_i)>1$ (or $<1$) indicates that compliers on average have a higher (or lower) infection probability than never takers under the control condition, given covariates. In other words, $\alpha_{i j}(\bcc_i)>1$ (or $<1$) implies that compliers are more susceptible (or healthier) than never takers. Similar interpretations extend to $\beta_{i j}(\bcc_i)$ and $\gamma_{i j}(\bcc_i)$. The goal of the sensitivity analysis is to explore how our estimates vary over a range of plausible values of the sensitivity functions. The following proposition shows that $\theta_g(a,a^*)$ is identified for fixed values of the sensitivity functions $\{\alpha_{i j}(\bcc_i),\beta_{i j}(\bcc_i),\gamma_{i j}(\bcc_i)\}$:
\begin{proposition}\label{prop:identification_pi}
Under Assumptions \ref{assum:population}--\ref{assum:monotonicity}, \ref{assum:positivity} and given sensitivity functions, we have
$$
\theta_g(a,a^*) = {\E\left[\frac{W_i}{N_i}\displaystyle\sum_{j=1}^{N_i} \omega_{i j}(g,a,a^*,\bm C_i)\times e_{i j}(g,\bm C_i) \times \mu_{i j}(a,d^*,\bm C_i)\right]}\Big/{\E\left[\frac{W_i}{N_i}\displaystyle\sum_{j=1}^{N_i} e_{i j}(g,\bm C_i)\right]},
$$
where $d^*$ is defined in Theorem \ref{thm:identification} and  $\omega_{i j}(g,a,a^*,\bm C_i)$ is defined as
\begin{align*}
& \omega_{i j}(\texttt{at}, 1,0,\bcc_i) = \frac{p_{i j}(1,1,\bcc_i)}{\beta_{i j}(\bcc_i)p_{i j}(1,1,\bcc_i)+(1-\beta_{i j}(\bcc_i))p_{i j}(0,1,\bcc_i)   }, \quad \omega_{i j}(\texttt{at}, 0,0,\bcc_i) = 1, \\
&   
 \omega_{i j}(\texttt{nt}, 0,0,\bcc_i) = \frac{p_{i j}(0,0,\bcc_i)}{1-\alpha_{i j}(\bcc_i) p_{i j}(0,1,\bcc_i)+(\alpha_{i j}(\bcc_i)-1)p_{i j}(1,1,\bcc_i) }, \quad \omega_{i j}(\texttt{nt},1,0,\bcc_i) = 1, \\
& \omega_{i j}(\texttt{co},1,1,\bcc_i) = \beta_{i j}(\bcc_i)\omega_{i j}(\texttt{at},1,0,\bcc_i),  \quad \omega_{i j}(\texttt{co},0,0,\bcc_i) = \alpha_{i j}(\bcc_i)\omega_{i j}(\texttt{nt},0,0,\bcc_i), \\
& \omega_{i j}(\texttt{co},1,0,\bcc_i) = \gamma_{i j}(\bcc_i). 
\end{align*}
\end{proposition}
Compared to Theorem \ref{thm:identification}, the identification formula in Proportion \ref{prop:identification_pi} leverages the weight function $\omega_{i j}(g,a,a^*,\bm C_i)$ to correct fir the bias due to principal non-ignorability. Proportion \ref{prop:identification_pi} simplifies to Theorem \ref{thm:identification} if all sensitivity functions are constantly 1. A moment estimator can then be constructed based on Proposition \ref{prop:identification_pi}, but it may be sensitive to working model misspecifications. Therefore, we continue to consider the EIF-induced estimators and extend the nonparametric efficient estimator $\widehat \theta_g^{\text{np}}(a,a^*)$ in Section \ref{sec:np} for sensitivity analysis. With a fixed value of the sensitivity functions, we derive the EIF of the bias-corrected identification formula in Theorem S1 in Web Appendix A.6. Based on the new EIF, the bias-corrected nonparametric efficient estimator of $\theta_g(a,a^*)$ has the following expression
\begin{equation*}
\widehat\theta_g^{\text{bc-np}}(a,a^*) = \frac{1}{K}\sum_{i=1}^K\frac{W_i}{N_i}\sum_{j=1}^{N_i}\widehat \psi_{i j}^{\text{bc},(1)}(a,d^*,a^{\dagger},d^{\dagger},h;\bm O_i)\Big/\frac{1}{K}\sum_{i=1}^K\frac{W_i}{N_i}\sum_{j=1}^{N_i}\widehat \psi_{ij}^{(2)}(a^{\dagger},d^{\dagger},h;\bm O_i).
\end{equation*}
Here, $\psi_{i j}^{\text{bc},(1)}(a,d^*,a^{\dagger},d^{\dagger},h;\bm O_i)$ is defined  
in Theorem S1 in Web Appendix A.6, which depends on $h_{\text{nuisance}}$ and sensitivity functions $\{\alpha_{ij}(\bcc_i),\beta_{ij}(\bcc_i),\gamma_{ij}(\bcc_i)\}$. Note that $\widehat\theta_g^{\text{bc-np}}(a,a^*)$ is still obtained via the cross-fitting procedure in Section \ref{sec:np}; i.e., all $\widehat \psi_{i j}^{\text{bc},(1)}(a,d^*,a^{\dagger},d^{\dagger},h;\bm O_i)$ and $\widehat \psi_{i j}^{(2)}(a^{\dagger},d^{\dagger},h;\bm O_i)$ for individuals among clusters in validation sample $\mathcal N_l$ are obtained with $h_{\text{nuisance}}$ estimated from machine learners based on clusters in the training sample $\mathcal N_{-l}$. The asymptotic behavior of $\widehat\theta_g^{\text{bc-np}}(a,a^*)$ is summarized below, which implies that a $o_p(K^{-1/4})$-type convergence rate for nuisance functions are sufficient for $\widehat\theta_g^{\text{bc-np}}(a,a^*)$ to be consistent, asymptotically normal, and semiparametrically efficient. 



{
\begin{theorem}\label{thm:bc-np}
Suppose that Assumptions \ref{assum:population}--\ref{assum:monotonicity}, \ref{assum:positivity} hold. Additionally, assume (a) conditions (a)--(b) in Theorem \ref{thm:np} hold; (b) for any $n_i\in \mathcal N$, $\bm v_i \in \mathcal V$, $\bm x_i \in \mathcal X(n_i)$, and $l\in\{1,\dots,L\}$, there exist $b_1>0$ such that $\omega_{ij}(g,a,a^*,\bm c_i)<b_1$ and $\widehat \omega_{ij,l}(g,a,a^*,\bm c_i)<b_1$, where $\widehat \omega_{ij,l}(g,a,a^*,\bm c_i)$ is the sensitivity weights evaluated under $\widehat p_{ij,l}(a,d,\bm c_i)$. Then, $\widehat \theta_{g}^{\text{bc-np}}(a,a^*)$ is consistent if $r_{p,K}=o(1)$, regardless of whether $r_{\mu,K}=o(1)$ or not. Furthermore, if $r_{p,K}=o(K^{-1/4})$ and $r_{\mu,K}=o(K^{-1/4})$, then $\sqrt{K}(\widehat \theta_{g}^{\text{bc-np}}(a,a^*)- \theta_{g}(a,a^*))$ converges to a zero-mean normal distribution with the asymptotic variance achieving the semiparametric efficiency bound defined in Theorem S1 in Web Appendix A.6. 
\end{theorem}
}

Based on $\widehat \theta_{g}^{\text{bc-np}}(a,a^*)$, one can further calculate $\widehat{\text{ICE}}_{g}^{\text{bc-np}}$, $\widehat{\text{NAE}}_{g}^{\text{bc-np}}$, or $\widehat{\text{PCE}}_{g}^{\text{bc-np}}$, across a grid of values of the sensitivity functions $\{\alpha_{i j}(\bcc_i),\beta_{i j}(\bcc_i),\gamma_{i j}(\bcc_i)\}$. {In Web Appendix~A.7, we provide data-guided recommendations for specifying plausible ranges of these sensitivity functions. Specifically, we show that the sensitivity functions have natural data-compatible bounds when the outcome is bounded, and that these ranges can be further calibrated using observed associations between baseline covariates and the outcome.} A detailed examination of the bias-corrected estimator shows that  $\widehat{\text{ICE}}_{g}^{\text{bc-np}}$, $\widehat{\text{NAE}}_{g}^{\text{bc-np}}$, and $\widehat{\text{PCE}}_{g}^{\text{bc-np}}$ only depend on one or two sensitivity functions, rather than all three simultaneously; the specific dependence for each estimand is summarized in Table \ref{tab:sensitivity}.  For purposes of inference, the asymptotic variance of $\widehat \theta_{g}^{\text{bc-np}}(a,a^*)$ (or $\widehat{\text{ICE}}_{g}^{\text{bc-np}}$, $\widehat{\text{NAE}}_{g}^{\text{bc-np}}$, and $\widehat{\text{PCE}}_{g}^{\text{bc-np}}$) can be estimated by the empirical variance of the corresponding EIFs. 

\begin{table}[h]
\caption{Dependence of the bias-corrected estimators on sensitivity functions 
$\alpha := \alpha_{ij}(\bcc_i)$, $\beta := \beta_{ij}(\bcc_i)$, and 
$\gamma := \gamma_{ij}(\bcc_i)$.\label{tab:sensitivity}}
\centering
\vspace{-0.3cm}
\scalebox{0.99}{ \begin{threeparttable}
\begin{tabular}{ccccccc}
\hline
\multirow{2}{*}{Principal Strata} & \multicolumn{3}{c}{Standard Monotonicity} & \multicolumn{3}{c}{Strong Monotonicity} \\ \cline{2-7} 
                                  & $\widehat{\text{PCE}}_{g}^{\text{bc-np}}$         & $\widehat{\text{ICE}}_{g}^{\text{bc-np}}$         & $\widehat{\text{NAE}}_{g}^{\text{bc-np}}$         & $\widehat{\text{PCE}}_{g}^{\text{bc-np}}$         & $\widehat{\text{ICE}}_{g}^{\text{bc-np}}$         & $\widehat{\text{NAE}}_{g}^{\text{bc-np}}$        \\ \hline
Compliers                         & $\{\alpha,\beta\}$   & $\{\beta,\gamma\}$   & $\{\alpha,\gamma\}$   & $\alpha$   & $\gamma$    & $\{\alpha,\gamma\}$   \\ \hline
Never-takers                      & $\alpha$   & None$^\mathparagraph$         & $\alpha$   & $\alpha$   & None$^\mathparagraph$          & $\alpha$   \\ \hline
Always-takers                     & $\beta$   & None$^\mathparagraph$           & $\beta$   & None$^\dagger$   & None$^\dagger$          & None$^\dagger$   \\ \hline
\end{tabular}
\begin{tablenotes}
      \item[$\mathparagraph$] Because $\text{ICE}_{\texttt{nt}} = \text{ICE}_{\texttt{at}}= 0$ by construction. 
      \item[$\dagger$] Because the always takers stratum does not exist under strong monotonicity. 
    \end{tablenotes}
\end{threeparttable}}
\end{table}

\vspace{-0.4in}

\section{Simulation studies}\label{sec:sim}

\subsection{Main simulation under extended principal ignorability}\label{sec:sim_main}

{We conduct simulations to evaluate the empirical performance of the proposed estimators, under the condition that Assumption \ref{assum:pi} holds. We simulate 1,000 CRTs, each containing $K=100$ clusters. For each cluster $i$, we first draw the complete data vector. Specifically, we first simulate $N_i$ from a discrete uniform distribution on $\{10,\dots,50\}$ and a univariate cluster-level covariate $V_i\sim N(\frac{3N_i}{50},1)$. Next, we generate individual-level covariate $X_{ij}\sim N(4,2^2)$. Under the standard monotonicity assumption, we generate the principal stratum $G_{ij} = \{D_{ij}(1),D_{ij}(0)\}$ given $\bcc_i = \{\bcx_i,V_i,N_i\}$ based on \eqref{eq:principal_score} with
$$
p_{ij}(a,d,\bcc_i) = \begin{cases}
\text{expit}\left(-8+ 4a + (1-a)N_i/50 + X_{ij} + V_i\right) , & \text{if } d=1,\\
1-\text{expit}\left(-8+ 4a + (1-a)N_i/50 + X_{ij} + V_i\right), & \text{if } d=0,
\end{cases} 
$$
following a logistic regression. Given $\{\bm D_i(1), \bm D_i(0), \bcx_i,V_i,N_i\}$, 
we generate the potential outcome $\{Y_{ij}(1),Y_{ij}(1,D_{ij}(0),\bm D_{i(-j)}(1)),Y_{ij}(0)\}$ for each individual based on the following linear structural model
\begin{align}
Y_{ij}(a,D_{ij}(a^*),\bm D_{i(-j)}(a)) = & 30 + \left(0.5 + \frac{3N_i}{100}+1.5X_{ij}\right)a + \left(0.2 + \frac{3N_i}{100}+1.5X_{ij}\right)D_{ij}(a^*) \nonumber \\
&  + \frac{1}{N_i-1}\sum_{j'\neq j} D_{ij'}(a) + X_{ij} + V_{i}  + \frac{N_i}{25} + \epsilon_{ij},  \label{eq:linear_structural_model_sim}
\end{align}
where the error term $\bm \epsilon_i = [\epsilon_{i1},\dots,\epsilon_{iN_i}]$ is drawn from a $N_i$-variate normal distribution with mean zero, variance $6^2$, and pairwise correlation 0.1. Assumption \ref{assum:pi} holds under this data-generating process because it aligns with the causal structural models outlined in Web Appendix A.1. Finally, we simulate $A_i\sim \text{Bernoulli}(0.5)$, and the observed data $\bco_i =\{\bm Y_i=\bm Y_i(A_i), \bm D_i=\bm D_i(A_i),A_i,\bcx_i,V_i,N_i\}$ is extracted from the the complete data.  

To focus ideas, we target the estimands $\text{ICE}_{\texttt{co}}$, $\text{NAE}_{\texttt{co}}$, $\text{NAE}_{\texttt{at}}$, and $\text{NAE}_{\texttt{nt}}$, where the cluster-specific weight is set to $W_i=1$, leading to cluster-average causal effects, with truth obtained via in a simulated super population with known potential outcomes.} For each estimand $\Delta$,  we evaluate the moment estimator ($\widehat \Delta^{\mathrm{mo}}$), doubly robust estimator ($\widehat \Delta^{\mathrm{dr}}$), and nonparametric efficient estimator ($\widehat \Delta^{\mathrm{np}}$). Here, $\widehat \Delta^{\mathrm{mo}}$ and $\widehat \Delta^{\mathrm{dr}}$ require specifying parametric models for $p_{ij}(a,d,\bm C_i)$ and  $\mu_{ij}(a,d,\bm C_i)$. To estimate $p_{ij}(a,d,\bm C_i)$, we consider a logistic regression for $D_{ij}$ on $\{A_{i},A_{i} \times N_i,X_{ij},V_{i},N_i\}$. To estimate $\mu_{ij}(a,d,\bm C_i)$, we consider a linear regression for $Y_{ij}$ on $\{D_{ij},D_{ij}\times N_i,D_{ij}\times X_{ij}, X_{ij},V_{i},N_i\}$ among clusters that randomized to arm $a$. {We show in Web Appendix A.8 that the parametric models align with the previous data-generating process approximately.}  
We consider a five-fold cross-fitting for $\widehat \Delta^{\mathrm{np}}$, and estimate the nuisance functions by Super Learner consisting of generalized linear model and random forest libraries \citep{phillips2023practical}; the input feature matrices were  identical to these in the corresponding parametric models.

We evaluate the performance of all three estimators under the following 4 scenarios depending on nuisance function specifications: (a) both $p_{ij}(a,d,\bm C_i)$ and $\mu_{ij}(a,d,\bm C_i)$ are correctly specified; (b) only $\mu_{ij}(a,d,\bm C_i)$ is correctly specified; (c) only $p_{ij}(a,d,\bm C_i)$ is correctly specified; (d) both $p_{ij}(a,d,\bm C_i)$ and $\mu_{ij}(a,d,\bm C_i)$ are misspecified. For correctly specified models, we specify the parametric model or Super Learner as previously mentioned. For misspecified models, we generate a set of transferred covariates $U_{ij}^{(1)}=\exp(-0.3X_{ij})$, $U_{ij}^{(2)}=V_i/(1+0.05X_{ij})$, and $U_{i}^{(3)}=(N_i\times V_i/25+0.6)^3$, and replace $\{X_{ij},V_i,N_i\}$ in the feature matrix of the parametric regression or Super Learner with $\{U_{ij}^{(1)},U_{ij}^{(2)},U_i^{(3)}\}$. While Super Learner offers greater flexibility than parametric working models, its performance may still depend on the quality of the input feature matrix, and the use of transformed variables in the feature matrix may introduce non-negligible bias for estimating the nuisance functions in finite-sample settings \citep{cheng2025inverting}.

\begin{figure}[ht]
\begin{center}
\includegraphics[width=1.0\textwidth]{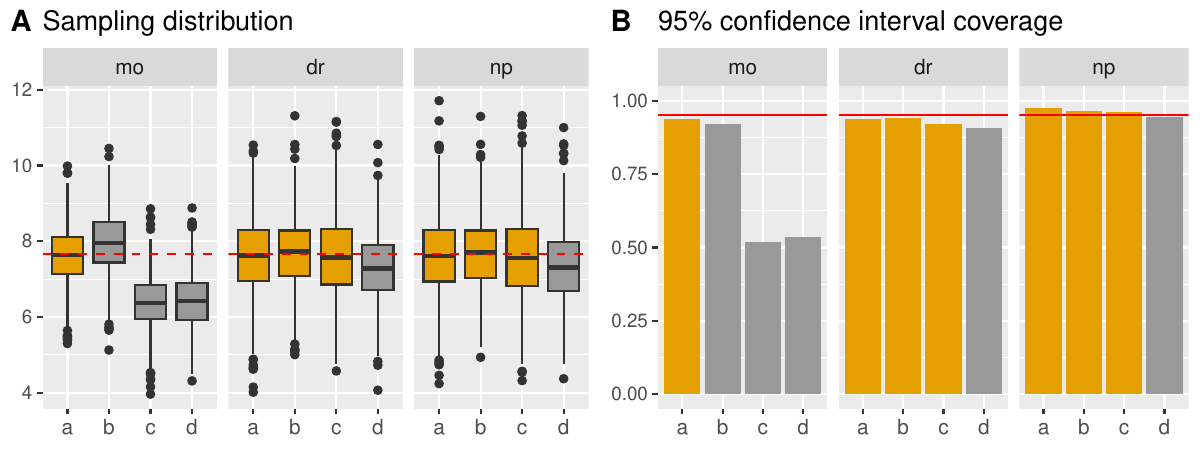}
\end{center}
\caption{Sampling distributions (Panel A) and 95\% Confidence Interval Coverage Probability (Panel B) for estimating $\text{NAE}_{\texttt{co}}$, based on the moment estimator (denoted by `mo'), doubly robust estimator (denoted by `dr'), and nonparametric efficient estimator (denoted by `np'), under scenarios (a)--(d). In Panel (A), the red dotted horizonal line indicates the true value of $\text{NAE}_{\texttt{co}}$.}
\label{fig:sim_NEEco_K100}
\end{figure}

\textcolor{black}{Figure \ref{fig:sim_NEEco_K100}(A) presents the sampling distributions for the three estimators of $\text{NAE}_{\texttt{co}}$ under Scenarios (a)--(d). Notably, $\widehat{\text{NAE}}_{\texttt{co}}^{\mathrm{mo}}$ is essentially unbiased in Scenario (a), but becomes biased in Scenarios (b)--(d) under model misspecification. $\widehat{\text{NAE}}_{\texttt{co}}^{\mathrm{dr}}$ provides additional protection against misspecification with negligible bias in Scenarios (a)--(c) when at most one of the parametric models is misspecified, but carries moderate bias in Scenario (d) when both models are misspecified. $\widehat{\text{NAE}}_{\texttt{co}}^{\mathrm{np}}$ shows similar performance to $\widehat{\text{NAE}}_{\texttt{co}}^{\mathrm{dr}}$, with negligible bias in Scenarios (a)--(c) and some bias in Scenario (d). We also investigate the 95\% confidence interval coverage rates in Figure \ref{fig:sim_NEEco_K100}(B). We construct 95\% confidence intervals based on bootstrapping for the moment-type and doubly robust estimators, and based on the proposed variance estimator for the nonparametric efficient approach. $\widehat{\text{NAE}}_{\texttt{co}}^{\mathrm{np}}$ achieves close-to-nominal coverage across all scenarios. $\widehat{\text{NAE}}_{\texttt{co}}^{\mathrm{dr}}$ attains nominal coverage in Scenarios (a)--(c), but exhibits slight under-coverage in Scenario (d), with coverage around 90\%.} 

\textcolor{black}{The performance for estimating $\text{ICE}_{\texttt{co}}$, $\text{NAE}_{\texttt{at}}$  and $\text{NAE}_{\texttt{nt}}$ is summarized in Web Figures 1--3 in the Supplementary Material, with qualitatively similar patterns. We also conduct simulations with $K=50$ clusters, and  results are provided in Web Figures 4--7 in the Supplementary Material for estimation of $\text{NAE}_{\texttt{co}}$, $\text{ICE}_{\texttt{co}}$, $\text{NAE}_{\texttt{at}}$ and $\text{NAE}_{\texttt{nt}}$, respectively. These simulation results are qualitatively similar to our main findings and and consistent with the prediction by our asymptotic analysis. Overall, the nonparametric efficient estimator performs slightly better than the doubly robust estimator in terms of both bias and coverage, especially in Scenario (d) when both nuisance models are misspecified.}

\subsection{Additional simulation under principal non-ignorability}

{We conduct simulations to evaluate the performance of the estimators when Assumption \ref{assum:pi} is violated. Specifically, we consider a setting in which $\bcx_i$ is unobserved, so that only $\bcc_i=\{V_i,N_i\}$ is included in the nuisance models for estimating $p_{ij}(a,d,\bcc_i)$ and $\mu_{ij}(a,d,\bcc_i)$. 
Because $X_{ij}$ is unobserved and associated with $G_{ij}$, omitting $X_{ij}$ from the linear structural model in \eqref{eq:linear_structural_model_sim} absorbs its contribution into the error term. The resulting composite error term is therefore correlated with $G_{ij}$, inducing dependence between $Y(a,D_{ij}(a^*),\bm D_{i(-j)}(a))$ and $G_{ij}$ conditioning only on $\bcc_i=\{V_i,N_i\}$, violating Assumption \ref{assum:pi}.

For each estimand $\Delta \in \{\text{ICE}_{\texttt{co}}, \text{NAE}_{\texttt{co}}, \text{NAE}_{\texttt{at}},  \text{NAE}_{\texttt{nt}}\}$, we evaluate the bias-corrected nonparametric efficient estimator $\widehat \Delta^{\mathrm{bc-np}}$. We also include the three estimators from Section \ref{sec:sim_main}, $\widehat \Delta^{\mathrm{mo}}$, $\widehat \Delta^{\mathrm{dr}}$, and $\widehat \Delta^{\mathrm{np}}$, as benchmarks; these estimators do not adjust for unmeasured confounding and are therefore expected to be biased. To implement $\widehat \Delta^{\mathrm{bc-np}}$, we need to specify the sensitivity functions $\{\alpha_{ij}(\bcc_i),\beta_{ij}(\bcc_i),\gamma_{ij}(\bcc_i)\}$. To simplify the analysis, we adopt a covariate-constant approximation under which these sensitivity functions do not depend on $\bcc_i$, reducing them to three sensitivity parameters $\{\alpha,\beta,\gamma\}$. We use a sufficiently large simulated data set with 10,000 clusters to approximate these sensitivity parameters. By definition, $\alpha$ represents the conditional risk ratio comparing compliers with never takers given $\bcc_i=\{V_i,N_i\}$. We therefore approximate $\alpha$ by fitting a log-binomial regression of $Y_{ij}(0)$ on $\{\mathbb{I}(G_{ij}=\texttt{co}),V_i,N_i\}$ among individuals who are either compliers or never takers in the large simulated data set. The exponentiated coefficient for $\mathbb{I}(G_{ij}=\texttt{co})$ is then used as the value of $\alpha$. The same procedure is used for calculating $\beta$ and $\gamma$, yielding $\alpha\approx 1.05$, $\beta \approx 0.87$, and $\gamma \approx 1.11$, which are then used to construct $\widehat \Delta^{\mathrm{bc-np}}$.

\begin{figure}[ht]
\begin{center}
\includegraphics[width=1.0\textwidth]{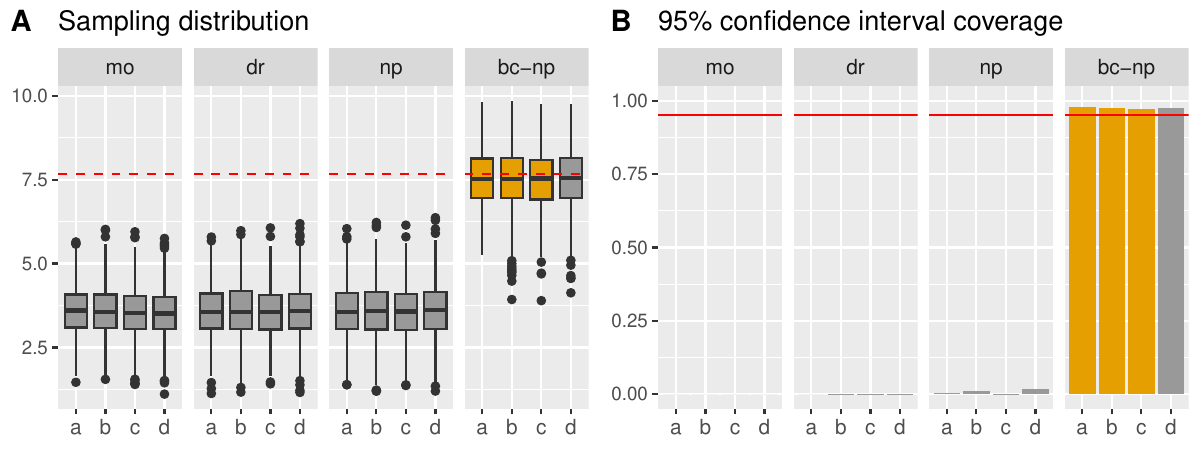}
\end{center}
\caption{Sampling distributions (Panel A) and 95\% Confidence Interval Coverage Probability (Panel B) for $\text{NAE}_{\texttt{co}}$ under scenarios (a)--(d), in the setting when Assumption \ref{assum:pi} is violated. In Panel (A), the red dotted horizonal line indicates the true value of $\text{NAE}_{\texttt{co}}$.}
\label{fig:sim_NEEco_sa}
\end{figure}

The specifications of the parametric and machine learning models follow those in Section \ref{sec:sim_main}, except that $X_{ij}$ is removed from all nuisance models. We again consider the four scenarios (a)--(d) as in Section \ref{sec:sim_main}, corresponding to different combinations of correct and incorrect nuisance model specification. For the misspecified models, we replace $\{V_i,N_i\}$ in the feature matrix of the parametric regressions or Super Learner with $\{U_i^{(1)}=\exp(-0.3V_i), U_i^{(2)}=\sqrt{N_i}\}$. Figure \ref{fig:sim_NEEco_sa} summarizes the performance for estimating $\text{NAE}_{\texttt{co}}$. As expected, across all Scenarios (a)--(d), $\widehat{\text{NAE}}_{\texttt{co}}^{\mathrm{mo}}$, $\widehat{\text{NAE}}_{\texttt{co}}^{\mathrm{dr}}$, and $\widehat{\text{NAE}}_{\texttt{co}}^{\mathrm{np}}$ exhibit substantial bias and close-to-zero coverage. In contrast, $\widehat{\text{NAE}}_{\texttt{co}}^{\mathrm{bc\text{-}np}}$ substantially reduces the bias and achieves conservative coverage across Scenarios (a)--(d). The performance for estimating $\text{ICE}_{\texttt{co}}$, $\text{NAE}_{\texttt{at}}$, and $\text{NAE}_{\texttt{nt}}$ is presented in Web Figures 8--10 in the Supplementary Material, with qualitatively similar results. To conclude, violations of the extended principal ignorability assumption can lead to severe bias for estimators that do not adjust for unmeasured confounding, whereas the proposed bias-corrected estimator can mitigate this bias when the sensitivity functions are correctly specified.} 

\vspace{-0.3in}

\section{Applications to the PSDP CRT}
\label{sec:app}

\subsection{Main analysis under extended principal ignorability}

We investigate the individual compliance effect and network assignment effect under the proposed causal framework for the PSDP CRT. Throughout, we set the cluster-specific weight to $W_i=1$, thereby targeting cluster-average causal effects. To support extended principal ignorability, we adjust for cluster-level covariates ($\bm V_i$), including school latrines per pupil, average exam score in 1996, geographic-zone infection prevalence in 1998, and distance to Lake Victoria, and individual-level covariates ($\bm X_{ij}$), including age, sex, weight-for-age Z-score, blood in stool, livestock and latrine ownership, recent malaria or fever, frequent illness, and observed pupil cleanliness. {As shown in Web Appendix A.9, we follow Section \ref{sec:sim} in specifying working models for $p_{ij}(a,d,\bm C_i)$ and $\mu_{ij}(a,d,\bm C_i)$, where $\bcc_i=\{\bcx_i,\bm V_i,N_i\}$. To assess the adequacy of the working models for the principal score $p_{ij}(a,d,\bcc_i)$, we developed covariate-balance diagnostics based on standardized mean differences \citep{ding2017principal,cheng2026multiply}; details are provided in Web Appendix A.10. In the PSDP study, these diagnostics show that the best balance is achieved by including $\min\{N_i,44\}$, rather than $N_i$, in both the parametric and machine-learning principal score models, where $44$ is the median cluster size. Under this specification, all components of $\{\bcx_i,\bm V_i\}$ are well balanced, although the balance diagnostic continues to indicate residual imbalance in $N_i$. Further inspection indicates substantial imbalance in $N_i$ between the treated and control groups\footnote{{In our analysis, $N_i$ refers to the number of pupils with measured 1999 parasitological outcomes, rather than school enrollment. $N_i$ is always smaller in treated schools ($A_i=1$, Group 1 schools in \cite{miguel2004worms}) than in control schools ($A_i=0$, Group 2 schools). As one limitation acknowledged in \cite{miguel2004worms}, this difference arises from the original data-collection design: Group 1 pupils were drawn from the 1998 parasitological sample and followed up in 1999 for outcome measurement, but some pupils were not resurveyed in 1999 because they were absent or had graduated by 1999. In contrast, Group 2 pupils were newly surveyed in 1999 for outcome measurement. We have also included an analysis of ICEs and NAEs excluding $N_i$ as a covariate, and the results are given in Web Appendix A.10.}}. We therefore retain the current model specification.}

Table \ref{tab:psdp_main} (Panel A) presents causal effects on the risk
ratio scale, including point estimates and 95\% confidence intervals
(CIs) for $\text{NAE}_{\texttt{co}}^{\mathrm{RR}}$ and
$\text{ICE}_{\texttt{co}}^{\mathrm{RR}}$ among compliers and
$\text{NAE}_{\texttt{nt}}^{\mathrm{RR}}$ among never takers. We also
calculate $\text{PCE}_{\texttt{co}}^{\mathrm{RR}}$ among compliers,
whereas $\text{PCE}_{\texttt{nt}}^{\mathrm{RR}}
=\text{NAE}_{\texttt{nt}}^{\mathrm{RR}}$
by construction. We first focus on the compliers. {
The nonparametric efficient estimator gives
$\widehat{\text{PCE}}_{\texttt{co}}^{\mathrm{RR},\text{np}}=0.589$
(95\% CI $[0.442,0.737]$), suggesting that assignment to deworming
reduces the risk of moderate-to-heavy worm infection by
approximately $41.1\%$. The nonparametric efficient estimator gives
$\widehat{\text{ICE}}_{\texttt{co}}^{\mathrm{RR},\text{np}}=0.761$
(95\% CI $[0.316,1.206]$), whose point estimate corresponds to an
approximately $23.9\%$ reduction in infection risk from individual
treatment uptake, but the CI includes $1$. Similarly, 
$\widehat{\text{NAE}}_{\texttt{co}}^{\mathrm{RR},\text{np}}=0.774$
(95\% CI $[0.329,1.220]$), whose point estimate suggests an
approximately $22.6\%$ reduction in risk from the network assignment
effect, although the CI also includes $1$. The moment and doubly robust estimators yield effects in the same protective direction for $\text{PCE}_{\texttt{co}}^{\mathrm{RR}}$ and $\text{ICE}_{\texttt{co}}^{\mathrm{RR}}$, while their estimates of $\text{NAE}_{\texttt{co}}^{\mathrm{RR}}$ are closer to the null. We next consider $\text{NAE}_{\texttt{nt}}^{\mathrm{RR}}$ among never takers. The nonparametric efficient estimator gives $\widehat{\text{NAE}}_{\texttt{nt}}^{\mathrm{RR},\text{np}}=0.803$
(95\% CI $[0.529,1.077]$), corresponding to an
approximately $19.7\%$ reduction in the risk of moderate-to-heavy worm
infection among never takers. Because
never takers do not take the medication themselves, any reduction may
reflect protective spillover from peers' uptake and/or other direct
assignment-related mechanisms. The moment and doubly robust estimators yield point estimates of $\text{NAE}_{\texttt{nt}}^{\mathrm{RR}}$ slightly above $1$, although neither differs significantly from the null. For completeness, we also report causal
effects on the risk difference scale in Table \ref{tab:psdp_main}
(Panel B), where the results are compatible with those on
the risk ratio scale.}

\begin{table}[ht!]
\caption{Analysis of the PSDP study, based on the moment estimator (denoted by `mo'), doubly robust estimator (denoted by `dr'), and nonparametric efficient estimator (denoted by `np'). The numbers in the brackets are the associated 95\% confidence interval.\label{tab:psdp_main}}
\centering
\renewcommand{\arraystretch}{1.3}
\scalebox{0.8}{ \begin{threeparttable}
\begin{tabular}{lcccc}
\hline
\multirow{2}{*}{Stratum} & \multirow{2}{*}{Estimand} & \multicolumn{3}{c}{Estimator} \\
\cline{3-5}
                    &  & \multicolumn{1}{c}{mo}                     & \multicolumn{1}{c}{dr}                     & \multicolumn{1}{c}{np}                      \\ \hline
\multicolumn{5}{c}{\textbf{Panel A: Risk ratio scale}}  \\
Compliers
    & $\text{NAE}_{\texttt{co}}^{\mathrm{RR}}$
    & 1.039 (0.621, 1.572)
    & 0.945 (0.347, 1.665)
    & 0.774 (0.329, 1.220) \\
    & $\text{ICE}_{\texttt{co}}^{\mathrm{RR}}$
    & 0.739 (0.518, 1.157)
    & 0.817 (0.472, 1.807)
    & 0.761 (0.316, 1.206) \\
    & $\text{PCE}_{\texttt{co}}^{\mathrm{RR}}$
    & 0.768 (0.503, 1.110)
    & 0.772 (0.508, 1.111)
    & 0.589 (0.442, 0.737) \\
\hline
Never Takers$^\mathparagraph$
    & $\text{NAE}_{\texttt{nt}}^{\mathrm{RR}}$
    & 1.081 (0.589, 1.740)
    & 1.074 (0.543, 1.763)
    & 0.803 (0.529, 1.077) \\
\hline
\multicolumn{5}{c}{\textbf{Panel B: Risk difference scale}}  \\
Compliers
    & $\text{NAE}_{\texttt{co}}$
    & 0.017 ($-$0.195, 0.192)
    & $-$0.023 ($-$0.314, 0.221)
    & $-$0.105 ($-$0.313, 0.104) \\
    & $\text{ICE}_{\texttt{co}}$
    & $-$0.116 ($-$0.254, 0.044)
    & $-$0.074 ($-$0.327, 0.149)
    & $-$0.086 ($-$0.292, 0.120) \\
    & $\text{PCE}_{\texttt{co}}$
    & $-$0.099 ($-$0.262, 0.035)
    & $-$0.097 ($-$0.264, 0.038)
    & $-$0.191 ($-$0.268, $-$0.114) \\
\hline
Never Takers$^\mathparagraph$
    & $\text{NAE}_{\texttt{nt}}$
    & 0.032 ($-$0.201, 0.235)
    & 0.029 ($-$0.204, 0.237)
    & $-$0.084 ($-$0.209, 0.041) \\ 
\hline

\end{tabular}
\begin{tablenotes}
      \item[$\mathparagraph$] For never takers, the network assignment effect equals the principal causal effect by construction.
    \end{tablenotes}
\end{threeparttable}}
\end{table}


\textcolor{black}{Under the network exclusion restriction (Remark \ref{remark:compare_to_spillover_efffect}), \cite{kang2019spillover} derived bounds on the network effects on the risk difference scale among compliers and never-takers in the PSDP CRT. Because the trial was unblinded, however, this restriction may be violated. That is, children assigned to treated schools could become more aware of the health risks of helminth infection and improve their hygiene, thereby affecting their outcomes even without taking the medication. Such behavioral responses constitute direct assignment effects. As clarified in Remark \ref{remark:compare_to_spillover_efffect}, if the network exclusion restriction fails to hold, the same bounds can instead be interpreted as bounds for $\text{PCE}_g$. We therefore compare these bounds with our point estimates of $\text{PCE}_g$ on the risk difference scale. Specifically, we adopt the covariate-assisted
approach of \cite{park2023assumption}, extending \cite{kang2019spillover} by using covariate-assisted classification to obtain sharp, potentially narrower bounds. After adjusting for the same set of covariates, we obtain that $\text{PCE}_{\texttt{co}}$ is bounded by $[-0.305,-0.175]$ with 95\% CI $[-0.445,-0.069]$, and $\text{PCE}_{\texttt{nt}}$ is bounded by
$[-0.590,0]$ with 95\% CI $[-0.694,0]$. Comparing to Table \ref{tab:psdp_main}, both analyses suggest a negative $\text{PCE}_{\texttt{co}}$, indicating that assignment
to the deworming intervention reduced infection risk for children in the
complier stratum. For never takers, their partial identification approach yields a wide bound, whereas our point identification analysis under Assumption \ref{assum:pi} provides a more efficient estimate. This comparison
illustrates the trade-off between the two approaches; the bounding analysis relies on weaker assumptions but may be less informative, while our framework leverages extended principal ignorability assumption for point identification and to further decompose $\text{PCE}_{\texttt{co}}$ into
$\text{ICE}_{\texttt{co}}$ and $\text{NAE}_{\texttt{co}}$.}

\subsection{Causal sensitivity analysis}

{Assumption~\ref{assum:pi} may be violated if unmeasured factors jointly affect both children's treatment uptake behavior and their subsequent infection outcomes. For example, school attendance around the time of treatment administration may influence whether students receive the deworming medication and may also be associated with later infection risk. Therefore, we examine the sensitivity of our results in the PSDP study under the violation of Assumption~\ref{assum:pi}.} Because strong monotonicity holds by the study design, we only need to specify the sensitivity functions $\{\alpha_{ij}(\bcc_i),\gamma_{ij}(\bcc_i)\}$ to study the sensitivity of the estimates among the compliers and never takers strata (Table \ref{tab:sensitivity}). Since there is no prior knowledge about how the sensitivity functions vary with the baseline covariates, we simplify by assuming that the sensitivity functions are constant across covariate levels; that is, we set $\alpha=\alpha_{ij}(\bcc_i)$ and $\gamma =\gamma_{ij}(\bcc_i)$. {Web Appendix~A.7 describes data-guided choices for the sensitivity parameter ranges. Specifically, we first use Proposition~S1 in the Supplementary Material to obtain broad data-compatible ranges implied by the bounded outcome condition $0\leq Y_{ij}\leq 1$, yielding $\alpha\in[0.42,\infty)$ and $\gamma\in[0,2.57]$. Because these ranges are wide, we further calibrate $\alpha$ and $\gamma$ using observed associations between baseline covariates and the potential outcome. Specifically, under the condition that the association between the principal stratum and the potential outcome is at most 1.5 times as large as the strongest observed covariate-outcome association, this calibration yields $\alpha\in[0.65,1.53]$ and $\gamma\in[0.65,1.53]$. By intersecting the calibration ranges and the data-compatible ranges, we consider 
$\alpha\in[0.65,1.53]$ and $\gamma\in[0.65,1.53]$ in our sensitivity analysis.}

\begin{figure}[ht]
\begin{center}
\includegraphics[width=0.8\textwidth]{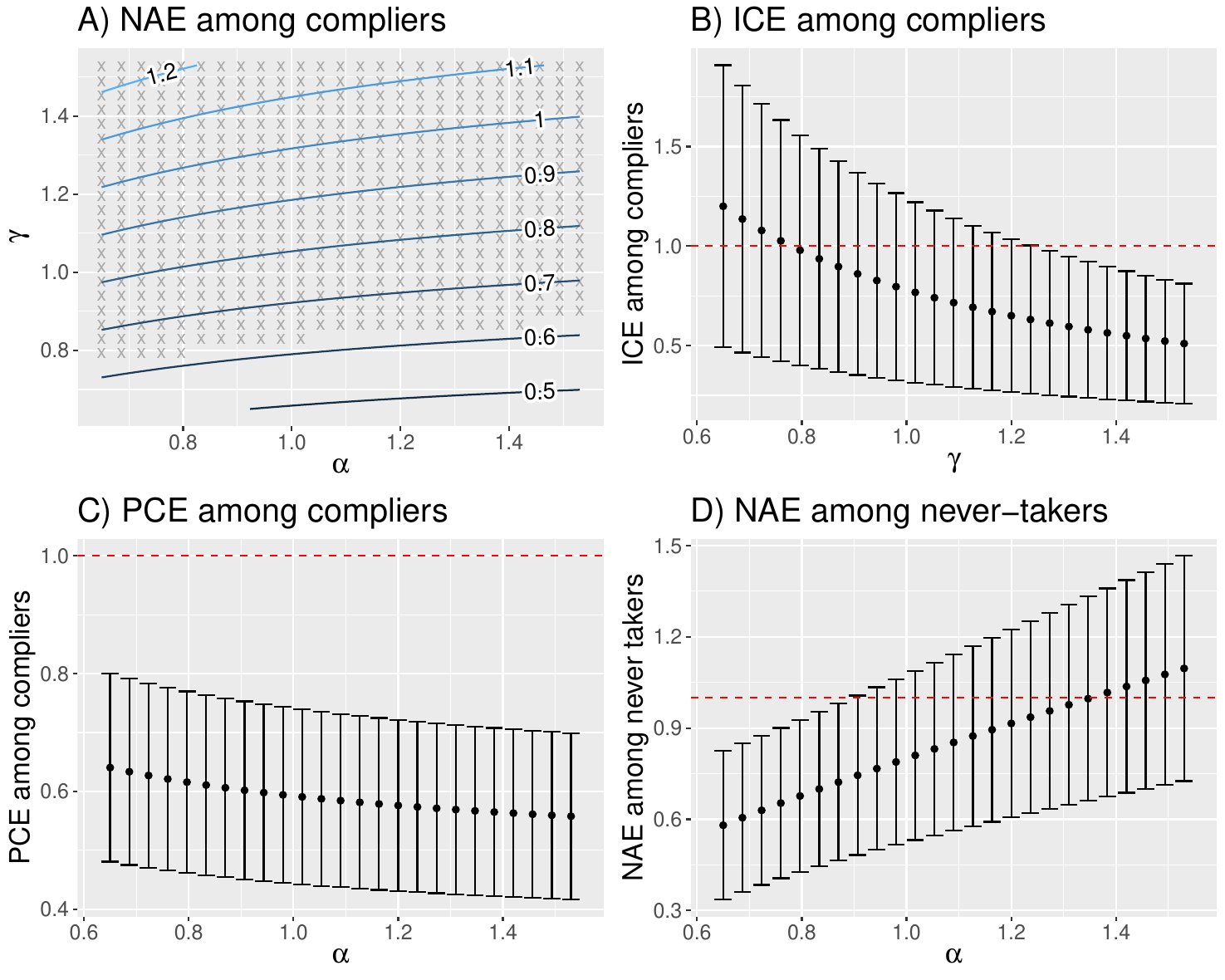}
\end{center}
\caption{Sensitivity analysis on violation of the extended principal ignorability assumption for the PSDP CRT, based on the bias-corrected nonparametric efficient estimator with $\{\alpha,\gamma\}$ varying within $[0.65,1.53]\times [0.65, 1.53]$. }
\label{fig:PSDP_SA}
\end{figure}

{Next, we estimate $\text{NAE}_{\texttt{co}}^{\mathrm{RR}}$, $\text{ICE}_{\texttt{co}}^{\mathrm{RR}}$, $\text{PCE}_{\texttt{co}}^{\mathrm{RR}}$, and $\text{NAE}_{\texttt{nt}}^{\mathrm{RR}}$ using the bias-corrected nonparametric efficient estimator and summarize the results in Figure \ref{fig:PSDP_SA}. By construction, $\widehat{\text{NAE}}_{\texttt{co}}^{\mathrm{RR},\text{bc-np}}$ depends on both $\alpha$ and $\gamma$, whereas $\widehat{\text{ICE}}_{\texttt{co}}^{\mathrm{RR},\text{bc-np}}$, $\widehat{\text{PCE}}_{\texttt{co}}^{\mathrm{RR},\text{bc-np}}$, and $\widehat{\text{NAE}}_{\texttt{nt}}^{\mathrm{RR},\text{bc-np}}$ depend only on $\gamma$, $\alpha$, and $\alpha$, respectively (Table \ref{tab:sensitivity}). Because the effects are defined on the risk ratio scale, $1$ represents the null effect, while values below (above) $1$ indicate protective (harmful) effects.  $\widehat{\text{NAE}}_{\texttt{co}}^{\mathrm{RR},\text{bc-np}}$ is sensitive to $\gamma$ but relatively robust to $\alpha$, where the point estimate exceeds $1$ when $\gamma>1.22$. Similarly, $\widehat{\text{ICE}}_{\texttt{co}}^{\mathrm{RR},\text{bc-np}}$ is sensitive to $\gamma$, with its point estimate exceeding $1$ when $\gamma < 0.80$. In contrast, $\widehat{\text{PCE}}_{\texttt{co}}^{\mathrm{RR},\text{bc-np}}$ and its 95\% confidence interval remain below $1$ across all considered values of $\alpha$, indicating greater robustness to violations of principal ignorability. For never takers, $\widehat{\text{NAE}}_{\texttt{nt}}^{\mathrm{RR},\text{bc-np}}$ increases with $\alpha$ and exceeds $1$ when $\alpha>1.37$. Overall, $\text{PCE}_{\texttt{co}}^{\mathrm{RR}}$ remains protective throughout the considered sensitivity range, whereas $\text{NAE}_{\texttt{co}}^{\mathrm{RR}}$ and $\text{ICE}_{\texttt{co}}^{\mathrm{RR}}$ are more sensitive to violations of principal ignorability. For never takers, the point estimate of $\text{NAE}_{\texttt{nt}}^{\mathrm{RR}}$ remains below $1$ when $\alpha<1.37$.}

\section{Discussion}

{The NAE combines two sources of effects: one arising from spillover generated by changes in peers' uptake within the same cluster, and the other arising from the individual's own direct effect of assignment, possibly through other unmeasured mechanisms. In the special case where the network exclusion restriction holds (Remark~\ref{remark:compare_to_spillover_efffect}), assignment does not exert any direct influence on individual outcomes other than through treatment receipt, and the NAE therefore reduces to a pure spillover effect. In general settings, however, assignment may induce additional behavioral or psychological responses that operate outside the uptake pathway, preventing the NAE from being interpreted solely as a spillover effect. This distinction also clarifies the nuisance modeling required by our framework. Because our target estimand includes the NAE, rather than a separately identified pure spillover effect, identification and estimation do not require specifying a full outcome regression that conditions on the entire vector of peers' treatment uptake, such as
$
\E[Y_{ij}\mid A_i,D_{ij},\bm D_{i(-j)},\bm C_i].
$
Instead, the NAE is defined as an assignment-based contrast within principal strata, where peers' treatment uptake changes naturally with the cluster-level assignment. This allows us to avoid directly modeling the potentially high-dimensional dependence of $Y_{ij}$ on $\bm D_{i(-j)}$. If the goal were to further decompose the NAE into a pure spillover component and a direct or placebo assignment component, then additional structural assumptions and nuisance functions involving peers' treatment uptake, such as the outcome regression above or a lower-dimensional exposure mapping for $\bm D_{i(-j)}$, would be needed. Future research may consider incorporating additional assumptions to enable the identification of these additional components and to advance the mechanistic understanding in complex CRTs.}

{Although Theorems \ref{thm:identification}--\ref{thm:bc-np} do not require parametric restrictions on the nuisance functions, implementation of the proposed estimators adopts a common nuisance function structure across individuals. Specifically, we estimate $p_{ij}(a,d,\bcc_i)$ and $\mu_{ij}(a,d,\bcc_i)$ using common prediction functions that are evaluated at individual-specific covariates and cluster-level covariates. The resulting predictions are therefore unit-specific through their covariate inputs, but the regression function itself is shared across individuals. This modeling strategy is natural in CRTs when individuals are considered exchangeable after conditioning on observed covariates. If individuals occupied certain ordered positions within clusters and the nuisance functions differed systematically across individuals, our estimators may be biased and future work is needed to alleviate conventional exchangeability modeling assumptions.}

{In our causal sensitivity analysis, we define sensitivity functions on the risk ratio scale. This choice is motivated by the PSDP application, in which the outcome is binary and risk ratios are directly interpretable and widely used in epidemiologic research \citep{tamhane2016prevalence}. On this scale, $1$ corresponds to equal risks across principal strata and hence no violation of principal ignorability; values above (below) $1$ indicate higher (lower) risk in one stratum than in the comparison stratum. However, the scale is not essential to the framework, which can also be formulated using risk differences and odds ratios. The scale may be chosen based on interpretability, investigator preference, and analytical convenience.}

\section*{Supplementary Material}

Data for the PSDP CRT is  available at \url{https://doi.org/10.7910/DVN/28038}. The Web Appendices are available on the first author’s website.

\singlespacing

\bibliographystyle{jasa3}
\bibliography{reference}

\end{document}